\documentclass{article}

\usepackage{PRIMEarxiv}

\usepackage[utf8]{inputenc} 
\usepackage[T1]{fontenc}    
\usepackage{hyperref}       
\usepackage{url}            
\usepackage{booktabs}       
\usepackage{amsfonts}       
\usepackage{amsmath}
\usepackage{nicefrac}       
\usepackage{microtype}      
\usepackage{lipsum}
\usepackage{graphicx}
\usepackage{subcaption}
\usepackage{mathpazo}
\usepackage{sectsty}
\usepackage{tikz}
\usepackage[capitalize]{cleveref}
\usepackage{subcaption}
\usepackage[shortlabels,inline]{enumitem}
\usepackage{float}

\title{Model-free system identification of surface ships in waves via Hankel dynamic mode decomposition with control}

\author{
  Giorgio Palma$^{a}$, Andrea Serani$^{a}$, Shawn Aram$^{b}$,
  David W. Wundrow$^{b}$, David Drazen$^{b}$, Matteo Diez$^{a,\star}$\\
  $^{a}$National Research Council-Institute of Marine Engineering, Rome, Italy\\
  $^{b}$Naval Surface Warfare Center-Carderock Division, West Bethesda, Maryland, U.S.A.\\
  $^\star$\texttt{matteo.diez@cnr.it} \\
}

\begin{document}

\begin{tikzpicture}[remember picture,overlay]
   \node [rectangle, fill=cyan, fill opacity=0.5, anchor=north, minimum width=\paperwidth, minimum height=3cm] at (current page.north) {};

   \node [anchor=north, minimum width=\paperwidth, minimum height=3cm, text width=\textwidth, align=center, text height=5ex, text depth=15ex, align=left] at (current page.north) {
     \sffamily\small
     \textbf{This is a preprint submitted to:} \textit{Ocean Engineering}
   };
\end{tikzpicture}

\maketitle

\begin{abstract}
This study introduces and compares the Hankel dynamic mode decomposition with control (Hankel-DMDc) and a novel Bayesian extension of Hankel-DMDc as model-free (i.e., data-driven and equation-free) approaches for system identification and prediction of free-running ship motions in irregular waves. 
The proposed DMDc methods create a reduced-order model using limited data from the system state and incoming wave elevation histories, with the latter and rudder angle serving as forcing inputs.
The inclusion of delayed states of the system as additional dimensions per the Hankel-DMDc improves the representation of the underlying non-linear dynamics of the system by DMD.
The approaches are statistically assessed using data from free-running simulations of a 5415M hull's course-keeping in irregular beam-quartering waves at sea state 7, a highly severe condition characterized by nonlinear responses near roll-resonance. The results demonstrate robust performance and remarkable computational efficiency.
The results indicate that the proposed methods effectively identify the dynamic system in analysis. 
Furthermore, the Bayesian formulation incorporates uncertainty quantification and enhances prediction accuracy. 
Ship motions are predicted with good agreement with test data over a 15 encounter waves observation window.
No significant accuracy degradation is noted along the test sequences, suggesting the method can support accurate and efficient maritime design and operational planning.
\end{abstract}


\section{Introduction}\label{s:intro}
Accurately predicting ship motions is crucial for the design, operation, and safety of maritime vessels. This is particularly true when considering seakeeping and maneuvering in adverse weather conditions. The availability of reliable predictive tools, along with the understanding
of the physics involved, is paramount for analyzing and developing robust and high-quality vessels that meet the International Maritime Organization (IMO) Guidelines and NATO Standardization Agreements (STANAG) for commercial and military ships, respectively, and for pushing the limits of their operational envelopes preserving safety.

The efficient and accurate prediction of ship responses in waves is a challenging problem, that can be addressed with various numerical tools and fidelity levels. 
Recent works \cite{stern2015, vanWalree2020, serani2021urans, aram2024cfd} demonstrated the ability of computational fluid dynamics (CFD) methods with unsteady Reynolds-averaged Navier-Stokes (URANS) formulations to assess ship performances in waves and extreme sea conditions.
Other studies employed potential flow solvers \cite{lin2006numerical,BelknapReed2019}, and also hybrid formulations where URANS and potential flow are combined to reduce computational costs and yet model important viscous-related features \cite{White2022}.
Although powerful, these analyses are typically computationally intensive. This is particularly true when the simulations aim to achieve the statistical convergence of relevant quantities of interest, possibly accounting for complex fluid-structure interactions with high-fidelity solvers.

A complementary approach consists of the data-driven surrogate modeling of the mentioned phenomena. In this perspective, system identification 
can be considered a methodology for developing efficient and accurate models, able to incorporate significant features from high-fidelity numerical tools, and experiments. 
System identification approaches can be classified as white, grey, or black box, based on the level of detail that shall be given by the user about the system to be modeled \cite{Miller2021}.
A white-box method typically requires the full knowledge of the ship parameters and the governing equations in some parametric form. Hydrodynamic coefficients, forces and moments, and other parameters are obtained from physical principles and semi-empirical formulae. 
When the information required for a white box approach is only partially available, \textit{e.g.,} only the structures of the relevant equations are known, a grey-box (also referred to as physics-informed) strategy can be used, and parameters can be extracted with various methods from simulated or measured data. 
In black-box modeling, only data about the motion time histories are available, and the identification method is used to create an input-output representation, which can also be referred to as equation-free approach.
From white to black box, the interpretability of the output models typically decreases, but the information required from the user is also reduced.

Several examples of grey-box system identification can be found in the literature. In \cite{Wang2015}, a 4-degree-of-freedom model for maneuvering motions of a surface vessel is combined with least square support vector machines to retrieve coefficients using data from nonlinear simulations. Support vector machines are used on experimental data from full-scale trials in \cite{LUO2014} for regression of the surge speed, sway speed, and yaw rate.
\cite{araki2012} used extended Kalman filtering and constrained least square method to estimate the maneuvering coefficients in a 6-degree-of-freedom mathematical model from data generated by various sources (experiments, CFD, and system-based computations). The extended Kalman filter is also central in \cite{Perera2015} in which the method is applied to a second-order nonlinear Nomoto model for vessel steering.
The inverse dynamics technique is used in \cite{ALEXANDERSSON2024} to estimate the forces acting on a ship during maneuvers from free-running model tests, investigating the generalization properties of the obtained models and adding different levels of physical information. 
\cite{Chillcce2023} developed a data-driven system identification approach to identify the hydrodynamic coefficients of a maneuvering model of a ship, using data from free-running tests and constraining the least-square estimation with physical limits to improve regression noise robustness.
An interesting approach is found in \cite{marlantes2024} where a hybrid machine learning method is combined with a physical low-fidelity model to predict ship motions in waves by correcting the model's forces. The hybridization allows for the use of a dataset of limited size to train the machine learning method and obtain good accuracy from the grey-box approach.

Machine learning, artificial neural networks, and deep learning algorithms have emerged in recent years as powerful surrogate modeling techniques when applied in black-box approaches, leveraging large datasets to train their models and achieve remarkable prediction accuracy. 
Following \cite{geffner2018model}, these approaches may be interpreted as model-free methods, in the sense that prior knowledge of the governing equations of the system is not required. In this regard, these approaches may be also referred to as data-driven and equation-free.
Some interesting works that use feedforward (FF) and recurrent neural networks (RNNs) can be found for ship motion prediction applications can be found in the literature. \cite{MOREIRA2003a} and \cite{Moreira2003b} applied RNNs to predict sway and yaw velocities in maneuvering from the knowledge of rudder range and ship's velocity.
LSTM are successfully used to learn nonlinear wave propagation and the nonlinear roll of a ship section in beam seas in \cite{XU2021}.
\cite{Diez2024} and \cite{delaguila2021} compared the accuracy of different architectures, such as FF, RNN, gated recurrent units networks (GRU), long-short term memory networks (LSTM), and bidirectional LSTM (biLSTM), in forecasting the motions of ships operating in waves from simulated data of measures on the field. \cite{Chen2024} introduced an adaptive cycle reservoir with regular jumps (CRJ) to enhance the capability of the network to adapt to new incoming data.
Convolutional neural networks are combined with LSTM in \cite{Lee2023}, using foreseen wave fields as inputs to predict the ship motions' time histories.  

However, the power of such deep learning methods is held back by the non-negligible training cost of the models, which can be particularly high when the data are obtained, for example, with high-fidelity solvers or experiments. While the reduction of the required training effort is an active research field, collecting sufficient data can be challenging, particularly for digital twinning and real-time applications in the operational environment. 

A promising model-free alternative consists in dynamic mode decomposition (DMD) \cite{schmid2010,kutz2016dynamic,mezic2021koopman}. 
The method is based on the Koopman operator theory, an alternative formulation of dynamical systems theory that provides a versatile framework for the data-driven and equation-free study of high-dimensional nonlinear systems \cite{mezic2017}. 
DMD builds a reduced-order linear model of a dynamical system, approximating the Koopman operator, from a small set of multidimensional input-output pairs, without requiring any specific knowledge and assumption about the system dynamics. 
The DMD operates equally on measured or simulated data and obtains the model from them with a direct procedure that, from a machine learning perspective, constitutes the training phase. 

Its data-driven nature, the non-iterative training, and the data-lean property contributed to the popularity of DMD as a reduced-order modeling technique 
and real-time forecasting algorithm
in several fields, such as fluid dynamics and aeroacoustics \cite{rowley2009, schmid2010, Tang2012, Semeraro2012, Song2013}, epidemiology \cite{Proctor2015}, neuroscience \cite{brunton2016}, finance \cite{mann2016}, etc.
The literature also includes several methodological extensions of the original DMD algorithm that have been presented to improve the accuracy of the decomposition and, more generally, extend the method's capabilities. 
Of particular interest to the present work are the Hankel-DMD \cite{mezic2017,Brunton2017,kamb2020time,mezic2021koopman,serani2023}, the DMD with control (DMDc) \cite{proctor2016dynamic, Proctor2018}, and their combination in the Hankel-DMD with control (Hankel-DMDc) \cite{Brunton2021,ZAWACKI2023}. 

The Hankel-DMD (also known as Augmented-DMD \cite{serani2023}, Time-Delay coordinates extended DMD \cite{Brunton2021}, and Time Delay DMD \cite{Dylewsky2022}) has been shown to be a powerful tool for improving the ability of the linear model to capture relevant features of nonlinear and chaotic dynamics extending the state of the system with time-delayed copies, as it has been demonstrated to yield the true Koopman eigenfunctions and eigenvalues in the limit of infinite-time observations \cite{mezic2017}. It has, for example, been successfully used in  \cite{mohan2018data} and \cite{Dylewsky2022} to predict the short-term evolution of electric loads on the grid.

The DMDc extends the DMD theory to the treatment of externally forced systems, enabling the separation of the free response from the input effect. 
Two representative studies of its applications are \cite{ALJIBOORY2024}, which synthesizes a novel real-time control technique for unmanned aerial quadrotors using DMDc enabling the control system to adapt promptly to environmental or system behavior change,
and \cite{dawson2015} in which DMDc has been applied to simulation data to obtain a reduced-order model (ROM) of the forces acting on a rapidly pitching airfoil.

In \cite{Brunton2021} the algorithmic variant combining control and state augmentation with time-delayed copies is called Time-Delay Coordinates DMDc, and it is introduced using the same number of delayed copies for the state and the input. 
Similarly, \cite{ZAWACKI2023} described the dynamic mode decomposition with input-delayed Control which, however, as the name suggests, includes time-delayed copies of the inputs only.

Restricting the focus to the naval field, interesting contributions on the use of DMD for the forecasting of ship trajectories, motions, and forces can be found in \cite{diez2022datadriven}, \cite{diez2022snh} where DMD is hybridized with an artificial neural network, and \cite{serani2023} that performed a statistical evaluation of the prediction performance of the DMD algorithm, including the use of state augmentation with derivatives and time delayed copies of the state. 
Furthermore, a comparative analysis between prediction algorithms based on DMD and different neural network architectures, such as standard and bidirectional long-short-term memory, gated recurrent unit, and feedforward neural networks, can be found in \cite{Diez2024}. 
These works focus on short-term real-time forecasting 
of a ship's coursekeeping in irregular waves, or performing a turning circle in regular waves, obtained by training the DMD model with time histories of the system's state from the immediate past. 
A similar approach is also used in \cite{Chen2023}, where DMD is applied to the prediction of rigid motions of a coursekeeping ship in regular waves.

The objective of this paper is to introduce the Hankel-DMDc and its novel Bayesian extension as effective and efficient approaches to model-free system identification of surface ships operating in waves, while discussing and comparing the results obtained by the two methods. 
Hankel-DMDc uses an independent number of delays for the system state and its input. 
The Bayesian extension includes uncertainty quantification, by considering Hankel-DMDc hyperparameters as stochastic variables.
The work aims to create an accurate ROM for the long-term prediction of ship motions, whose efficiency is provided by the small dimensionality of the relevant state variables and the low computational cost required both for the model construction (training) and exploitation (prediction), as opposed to more data and resource-intensive machine learning methods.

Results are presented for the course keeping of the 5415M hull, free-running in irregular beam-quartering waves at Fr = 0.33 and sea state 7. This operating condition represents a highly severe scenario, characterized by nonlinear responses, making it a particularly challenging test case. 
\begin{figure}[ht!]
\centering
\includegraphics[width=0.75\linewidth]{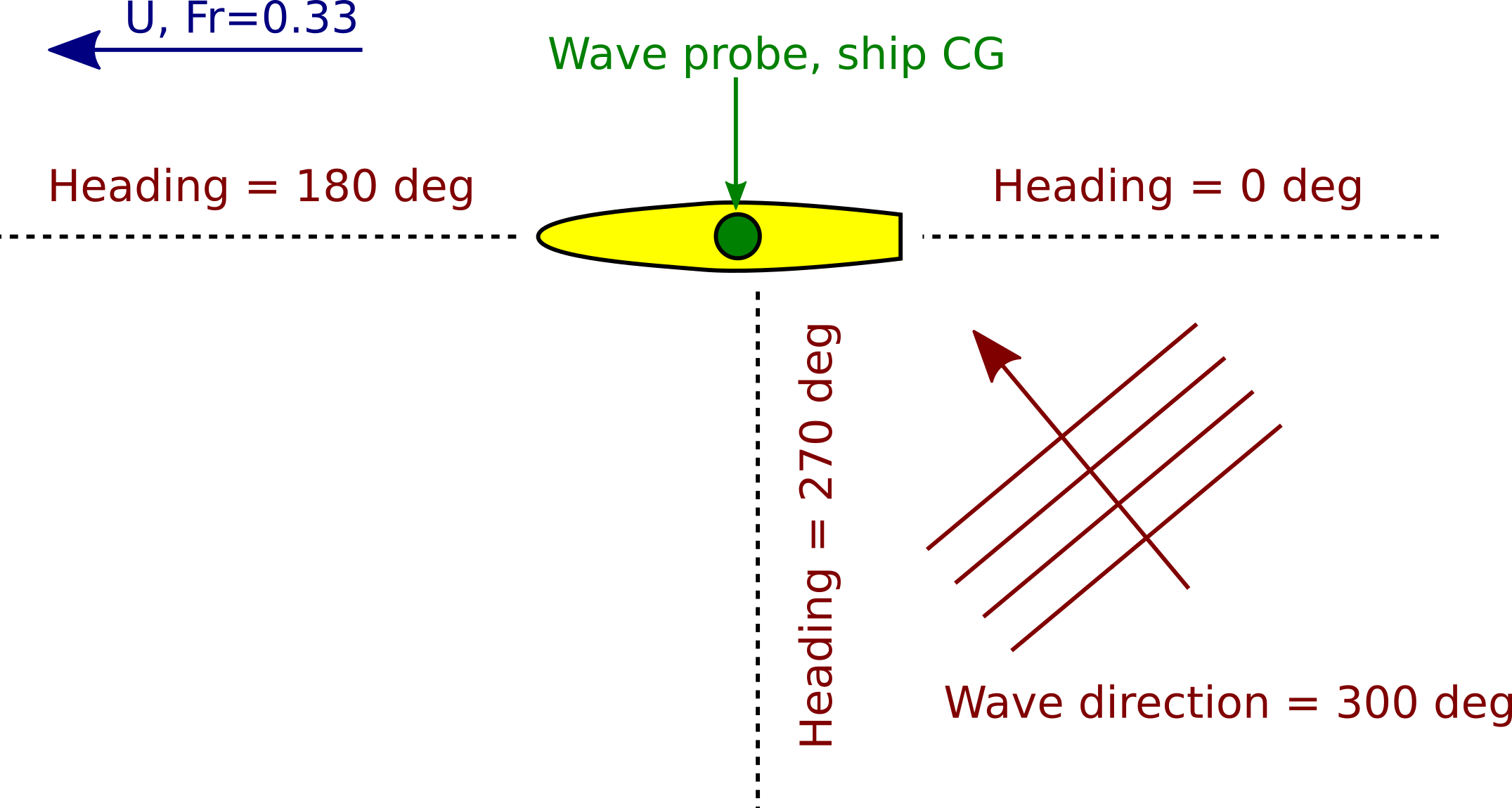}
  \caption{Setup and notation for the current test case}
  \label{fig:scheme}
  \bigskip
\end{figure}
The most promising hyperparameter sets for Hankel-DMDc are identified after testing a full-factorial combination of setting parameters against a set of randomly extracted validation sequences from the data record, using three evaluation metrics. The Bayesian Hankel-DMDc setup is determined by exploiting the findings of the deterministic design-of-experiment and tested against the same test sequences of the deterministic Hankel-DMDc for comparison.

The proposed methodology may pave the way to significant advancements in the design and optimization of vessels operating in waves, reducing the burden associated with simulations and computational efforts, in particular in situations where statistical convergence of quantity of interest is sought, consequently enhancing structural health monitoring and improving operational planning and safety. It can also drive further research on the development of more accurate ship simulators, incorporating realistic motion predictions in real-time, leading to safer, more efficient, and sustainable maritime operations.

The rest of the paper is structured as follows. \Cref{s:dmdc} introduces the DMDc formulation; its extension to Hankel-DMDc is described in \cref{s:hdmdc}; finally, the novel Bayesian Hankel-DMDc formulation is presented in \cref{s:bhdmdc}.
\cref{s:metrics} presents the three error metrics used in the current study to assess the performance of the DMD-based models.
\cref{s:test} describes the test case to which the methodology is applied.
The numerical setup for the DMD algorithms is defined in \cref{s:nums}.
The results are presented and discussed in \cref{s:res}, statistically assessing the prediction performances of the algorithms, and showing comparisons between the reference and predicted time histories for some representative test sequences.
Finally, concluding remarks can be found in \cref{s:concl}.

\section{Dynamic mode decomposition}\label{s:meth}
Dynamic mode decomposition was originally presented in \cite{schmid2008dynamic} and \cite{schmid2010} as a technique to analyze time series data from dynamical systems, decomposing them into a set of dominant modes that express the dominant dynamic behavior captured in the data sequence, and potentially predict their future temporal behavior. 
The method is an equation-free data-driven approach that has been shown by \cite{rowley2009} to be a computation of the Koopman operator for linear observables \cite{Marusic2024}.
The Koopman operator theory is built upon the original work by \cite{Koopman1931}, defining the possibility of transforming a non-linear dynamical system into a possibly infinite-dimensional linear system \cite{Proctor2018}. 
DMD, hence, can be considered a form of system identification, able to identify a linear ROM through its state transition matrix.

The DMD algorithm has been generalized and improved since its introduction and its state-of-the-art definition has been given by \cite{Tu2014}.
Several variants of the base algorithm have been presented in the literature, extending the algorithm capability \cite{kutz2016dynamic}. 
In particular, in the present work, the focus is on the DMD with control (DMDc) and in the Hankel-DMD with control algorithms. 

\subsection{Dynamic mode decomposition with control}\label{s:dmdc}
The original DMD algorithm characterizes naturally evolving dynamical systems. The DMDc aims at extending the algorithm to include in the analysis the influence of forcing inputs and disambiguate it from the unforced dynamics of the system \cite{proctor2016dynamic}. 
This algorithm extension is deemed suitable for the identification of a ROM for the ship's motion in waves, as the system evolution is influenced by inputs such as the elevation of waves approaching the vessel, and the actuation of the rudder.

A forced dynamic system in its discrete-time version is expressed as:
\begin{equation} \label{eq:discdynsys}
    \mathbf{x}_{j+1} = \mathrm{\mathbf{F}}(\mathbf{x}_j,j,\gamma,\mathbf{u_j}).
\end{equation}
where $\mathbf{F}$ is the (possibly) non-linear mapping describing the evolution of the state $\mathbf{x} \in \mathbb{R}^n$ from the time step $j$ to $j+1$, subject to the forcing $\mathbf{u}  \in \mathbb{R}^l$, for any discrete time $t_j=j\,\Delta t$, with $\gamma$ describing the parameters of the system.

DMDc treats the system $\mathbf{F}(\mathbf{x}_j,j,\gamma,\mathbf{u_j})$ as an unknown, \textit{i.e.}, without any assumptions of underlying physics, 
and applies an equation-free approach to develop a locally linear model of the system \cite{kutz2016dynamic} based on observed data, described by:
\begin{equation}\label{eq:dmdcdsys}
    \mathbf{x}_{j+1} = \mathrm{\mathbf{A}}\mathbf{x}_j + \mathbf{B}\mathbf{u}_j
\end{equation}
where $\mathbf{B} \in \mathbb{R}^{n \times l}$, and the discrete-time matrix $\mathbf{A} \in \mathbb{R}^{n\times n}$ is related to its homologous for continuous time by $\mathrm{\mathbf{A}}=\text{exp}(\mathcal{A}\Delta t)$.

Introducing the vector $\mathbf{y}_j$
\begin{equation}\label{eq:Y}
\mathbf{y}_j=
\begin{bmatrix}
\mathbf{x}_j \\
\mathbf{u}_j\\
\end{bmatrix},
\end{equation}
\cref{eq:dmdcdsys} can be rewritten as:
\begin{equation}\label{eq:dmdSIY}
\mathbf{x}_{j+1}=\mathbf{Gy}_j, \hspace{1cm} \text{with} \hspace{1cm} \mathbf{G}=
\begin{bmatrix}
    \mathbf{A} & \mathbf{B} \\
\end{bmatrix}.
\end{equation}
Collecting a set of $m$ snapshots of the system's state and input at different time steps $j=1,\dots,m$, the matrices $\mathbf{A}$ and $\mathbf{B}$ are obtained as the ones providing the best fitting of the sampled data in least squares sense. To this aim, the collected data are arranged in the following matrices:
\begin{equation}\nonumber
\mathbf{Y}=
\begin{bmatrix}
\mathbf{y}_j & \mathbf{y}_{j+1} & \dots & \mathbf{y}_{m-1}\\
\end{bmatrix},
\end{equation}
\begin{equation}\label{eq:XX'}
\mathbf{X}'=
\begin{bmatrix}
\mathbf{x}_{j+1} & \mathbf{x}_{j+2} & \dots & \mathbf{x}_{m}\\
\end{bmatrix},
\end{equation}
and the DMD approximation of the matrix $\mathbf{G}$ can be obtained from
\begin{equation}\label{eq:approxG}
\mathbf{G}\approx\mathbf{X}'\mathbf{Y}^{\dag},
\end{equation}
where $\mathbf{Y}^{\dag}$ is the Moore-Penrose pseudo-inverse of $\mathbf{Y}$, which minimizes $\|\mathbf{X}'-\mathbf{GY}\|_F$, where $\|\cdot\|_F$ is the Frobenius norm. 

The pseudoinverse of $\mathbf{Y}$ can be efficiently evaluated using the singular value decomposition (SVD)
\begin{equation}\label{eq:gsvd}
    \mathbf{G} = \mathbf{X}' \mathbf{V} \boldsymbol{\Sigma}^{-1}\mathbf{U}^*.
\end{equation}
The DMD approximations of the matrices $\mathbf{A}$ and $\mathbf{B}$ are obtained by splitting the operator $\mathbf{U}$ in $\mathbf{U_1} \in \mathbb{R}^{n \times n}$ and $\mathbf{U_2} \in \mathbb{R}^{l\times n}$ \footnote{Due to the low dimensionality of data in the current context, the pseudoinverse is computed using the full SVD decomposition with no rank truncation. Otherwise, truncating the SVD to rank $p$ one would have $\mathbf{U_1} \in \mathbb{R}^{n\times p}$ and $\mathbf{U_2} \in \mathbb{R}^{l\times p}$.}. 
\begin{equation}\label{eq:abdmdc}
    \mathbf{A} = \mathbf{X'}\mathbf{V}\boldsymbol{\Sigma}^{-1}\mathbf{U_1^*}, \hspace{1cm} \mathbf{B} = \mathbf{X'}\mathbf{V}\boldsymbol{\Sigma}^{-1}\mathbf{U^*_2}.
\end{equation}
%


The system dynamics are subsequently solved through Eq. \ref{eq:dmdcdsys}, 
with given initial conditions and the sequence of inputs $\mathbf{u}_j$.

\subsection{Hankel dynamic mode decomposition with control}\label{s:hdmdc}
The standard DMDc formulation approximates the Koopman operator in a restricted space of linear measurements, creating a best-fit linear model linking sequential data snapshots \cite{schmid2010,kutz2016dynamic}. 
This linear DMD provides a locally linear representation of the dynamics that can't capture many essential features of nonlinear systems.
The augmentation of the system state is thus the subject of several DMD algorithmic variants \cite{otto2019,takeishi2017,Lusch2018,Brunton2021}, aiming to find a coordinate system (or \textit{embedding}) that spans a Koopman-invariant subspace, in order to search for an approximation of the Koopman operator valid also far from fixed points and periodic orbits in a larger space.
The need for state augmentation through additional observables is even more critical for applications in which the number of states in the system is small, typically smaller than the number of available snapshots.
However, there is no general rule for defining these observables and guaranteeing they will form a closed subspace under the Koopman operator \cite{brunton2016b}.

The Hankel-DMD \cite{mezic2017} is a specific version of the DMD algorithm that has been developed to deal with the cases of nonlinear systems in which only partial observations are available such that there are \textit{latent} variables \cite{Brunton2021}.
The state vector is thus augmented, embedding time-delayed copies of the original variables, resulting in an intrinsic coordinate system that forms an invariant subspace of the Koopman operator (the time-delays form a set of observable functions that span a finite-dimensional subspace of Hilbert space, in which the Koopman operator preserves the structure of the system \cite{Brunton2017, Pan2020}).
The Hankel augmented DMD, hence, can better represent the underlying nonlinear dynamics of the systems, capturing their important features.

The formulation of the Hankel-DMD with control (Hankel-DMDc) can be obtained from the DMDc presented in \cref{s:dmdc} by transforming the matrices $\mathbf{Y}$ and $\mathbf{X}'$ 
in $\widehat{\mathbf{Y}}$ and  $\widehat{\mathbf{X}}'$, respectively.
State augmentation is achieved by adding a number $s$ of time-shifted copies of the time histories of the original states (delayed states) and $z$ time-shifted copies of the inputs (delayed inputs) to the data matrices:
\begin{equation}\label{eq:sXX'}
\widehat{\mathbf{Y}}=
\begin{bmatrix}
\mathbf{X} \\
\mathbf{S}\\
\mathbf{U} \\
\mathbf{Z}\\
\end{bmatrix},
\qquad
\widehat{\mathbf{X}}'=
\begin{bmatrix}
\mathbf{X}' \\ 
\mathbf{S}'\\
\end{bmatrix}.
\end{equation}
where the Hankel matrices $\mathbf{S}$, $\mathbf{S}'$, and $\mathbf{Z}$ are:  
\begin{equation}\label{eq:sts'}
\begin{split}
\mathbf{S}&=
\begin{bmatrix}
\mathbf{x}_{j-1} & \mathbf{x}_{j} & \dots & \mathbf{x}_{m-2}\\
\mathbf{x}_{j-2} & \mathbf{x}_{j-1} & \dots & \mathbf{x}_{m-3}\\
\vdots & \vdots & \vdots & \vdots \\
\mathbf{x}_{j-s-1} & \mathbf{x}_{j-s} & \dots & \mathbf{x}_{m-s-1}\\
\end{bmatrix}, \quad
\mathbf{S}'=
\begin{bmatrix}
\mathbf{x}_{j} & \mathbf{x}_{j+1} & \dots & \mathbf{x}_{m-1}\\
\mathbf{x}_{j-1} & \mathbf{x}_{j} & \dots & \mathbf{x}_{m-2}\\
\vdots & \vdots & \vdots & \vdots \\
\mathbf{x}_{j-s} & \mathbf{x}_{j-s+1} & \dots & \mathbf{x}_{m-s}\\
\end{bmatrix}, \\
\mathbf{Z}&=
\begin{bmatrix}
\mathbf{u}_{j-1} & \mathbf{u}_{j} & \dots & \mathbf{u}_{m-2}\\
\mathbf{u}_{j-2} & \mathbf{u}_{j-1} & \dots & \mathbf{u}_{m-3}\\
\vdots & \vdots & \vdots & \vdots \\
\mathbf{u}_{j-z-1} & \mathbf{u}_{j-z} & \dots & \mathbf{u}_{m-z-1}\\
\end{bmatrix}.    
\end{split}
\end{equation}
The algorithm is otherwise unaltered, and the definition of the system matrices $\mathbf{A}$ and $\mathbf{B}$ follows Eqs.(\ref{eq:approxG})-(\ref{eq:abdmdc}) using the augmented data matrices.
Once those are obtained, the evolution of the system follows a modified version of \cref{eq:dmdcdsys}, including the delayed state and input:
\begin{equation}\label{eq:hdmdcdsys}
    \mathbf{\hat{x}}_{j+1} = \mathrm{\mathbf{A}}\mathbf{\hat{x}}_j + \mathbf{B}\mathbf{\hat{u}}_j
\end{equation}
where $\hat{\mathbf{x}}_j = [\mathbf{x}_j,\,\mathbf{x}_{j-1},\,\dots\,,\mathbf{x}_{j-s}]^T$ and $\hat{\mathbf{u}}_j = [\mathbf{u}_j,\,\mathbf{u}_{j-1},\,\dots\,,\mathbf{u}_{j-z}]^T$
are the extended state and input vectors, including the respective $s$ and $z$ delayed copies.

\subsection{Bayesian extension of Hankel dynamic mode decomposition with control}\label{s:bhdmdc}
The shape and the values within matrices $\mathbf{A} \in \mathbb{R}^{ns \times ns}$ and $\mathbf{B} \in \mathbb{R}^{ns \times lz}$ produced by the Hankel-DMDc depend on the hyperparameters of the algorithm, such as the observation time length, $l_{tr} = t_m - t_1$, the maximum delay time in the augmented state $l_{d_x} = t_{j-1} - t_{j-s-1}$ and in the augmented input  $l_{d_u} = t_{j-1} - t_{j-z-1}$.
These dependencies can be denoted as follows:
\begin{equation}\label{eq:bayes1}
\mathbf{A}=\mathbf{A}(l_{tr},l_{d_x},l_{d_u}), \qquad \mathbf{B}=\mathbf{B}(l_{tr},l_{d_x},l_{d_u}).
\end{equation}
In the Bayesian Hankel-DMDc formulation, the three hyperparameters are considered stochastic variables, introducing uncertainty in the process.
Through uncertainty propagation, the solution $\mathbf{x}(t)$ also depends on $l_{tr}$,$l_{d_x}$ and $l_{d_u}$:
\begin{equation}\label{eq:bayes2}
\mathbf{x}(t)=\mathbf{x}(t,l_{tr},l_{d_x},l_{d_u}),
\end{equation}
and the following are used to define, at a given time $t$, the expected value of the solution and its standard deviation:
\begin{equation}\label{eq:bayes3}
\mathbf{\mu_x}(t)=\int_{l_{d_u}^l}^{l_{d_u}^u} \int_{l_{d_x}^l}^{l_{d_x}^u} \int_{l_{tr}^l}^{l_{tr}^u}\mathbf{x}(t,l_{tr},l_{d_x},l_{d_u})p(l_{tr})p(l_{d_x})p(l_{d_u})dl_{tr} dl_{d_x} dl_{d_u},
\end{equation}
\begin{equation}\label{eq:bayes4}
\mathbf{\sigma_x}(t)= \left\{ \int_{l_{d_u}^l}^{l_{d_u}^u} \int_{l_{d_x}^l}^{l_{d_x}^u} 
 \int_{l_{tr}^l}^{l_{tr}^u} \left[\mathbf{x}(t,l_{tr},l_{d_x})-\mathbf{\mu_x}(t) \right]^2 p(l_{tr})p(l_{d_x})p(l_{d_u})dl_{tr} dl_{d_x} dl_{d_u}\right\}^\frac{1}{2},
\end{equation}

where ${l_{tr}^l}$, ${l_{d_x}^l}$, ${l_{d_u}^l}$ and ${l_{tr}^u}$, ${l_{d_x}^u}$, ${l_{d_u}^u}$ are lower and upper bounds and $p(l_{tr})$, $p(l_{d_x})$, $p(l_{d_u})$, are the given probability density functions for $l_{tr}$, $l_{d_x}$ and $l_{d_u}$.

In practice, a uniform probability density function is assigned to the hyperparameters and a set of realizations is obtained through a Monte Carlo sampling. Accordingly, for each realization of the hyperparameters, the solution  $\mathbf{x}(t,l_{tr},l_{d_x},l_{d_u})$ is computed, and at a given time $t$ the expected value and standard deviation of the solution are then evaluated.

\section{Performance metrics}\label{s:metrics}
To evaluate the predictions made by the models and to compare the effectiveness of different methodologies and configurations, three error indices are employed: the normalized mean square error (NRMSE) \cite{Diez2024}, the normalized average minimum/maximum absolute error (NAMMAE) \cite{Diez2024}, and the Jensen-Shannon divergence (JSD) \cite{marlantes2024}. 
All the metrics are averaged over the variables that constitute the system's state. This comprehensive evaluation considers aspects such as overall error, the range, and the statistical similarity of predicted versus measured values.

The NRMSE quantifies the average root mean square error between the predicted values $\mathrm{\mathbf{\tilde x}}_t$ and the measured (reference) values $\mathrm{\mathbf{x}}_t$ at different time steps. 
It is calculated by taking the square root of the average squared differences, normalized by eight times the standard deviation of the measured values:

\begin{equation}\label{eq:nrmse}
   \mathrm{NRMSE} = \frac{1}{N} \sum_{i=1}^{N} \sqrt{\frac{1}{\mathcal{T} (8\sigma_{x_i})^2} \sum_{j=1}^{\mathcal{T}} \left( \tilde{x}_{ij} - x_{ij} \right)^2},
\end{equation}
where $N$ is the number of variables in the predicted state, $\mathcal{T}$ is the number of considered time instants, and $\sigma_{x_i}$ is the standard deviation of the measured values in the considered time window for the variable $x_i$.

The NAMMAE metric provides an engineering-oriented assessment of the prediction accuracy. It measures the absolute difference between the minimum and maximum values of the predicted and measured time series, as follows:
\begin{equation}
    \mathrm{NAMMAE} = \frac{1}{2 N (8\sigma_{x_i})} \sum_{i=1}^{N} \left( \left| \min_j(\tilde{x}_{ij}) - \min_j(x_{ij}) \right| + \left| \max_j(\tilde{x}_{ij}) - \max_j(x_{ij}) \right| \right),
\end{equation}

Lastly, the JSD provides a measure of the similarity between the probability distribution of the predicted and reference signal \cite{marlantes2024}.
For each variable, it estimates the entropy of the predicted time series probability density function $Q$ relative to the probability density function of the measured time series $R$, where $M$ is the average of the two \cite{Lin1991}. 
\begin{align}
    &\mathrm{JSD} = \frac{1}{N} \sum_{i=1}^{N} \left( \frac{1}{2}D(Q_i\,||\,M_i) + \frac{1}{2}D(R_i\,||\,M_i) \right),  \quad \\
    &\quad \text{with} \quad M = \frac{1}{2} (Q + R)\label{eq:jsd}, \\
    &\quad \text{and} \quad D(K\,||\,H)=\sum_{y \in \chi} K(y) \ln\left( \frac{K(y)}{H(y)} \right). \label{eq:kld}
\end{align}
The Jensen-Shannon divergence is based on the Kullback-Leibler divergence $D$, given by \cref{eq:kld}, which is 
the expectation of the logarithmic difference between the probabilities $K$ and $H$, both defined over the domain $\chi$, where the expectation is taken using the probabilities $K$ \cite{Kullback1951}
The similarity between the distributions is higher when the Jensen-Shannon distance is closer to zero. JSD is upper bounded by $\ln(2)$.

Each of the three indices contributes to the error assessment with its peculiarity, providing a holistic assessment of prediction accuracy:
\begin{itemize}
    \item[-] The NRMSE evidences phase, frequency, and amplitude errors between the reference and the predicted signal, evaluating a pointwise difference between the two. However, it is not possible to discern between the three types of error and to what extent each type contributes to the overall value.
    \item[-] The NAMMAE indicates if the prediction varies in the same range of the original signal, but does not give any hint about the phase or frequency similarity of the two.
    \item[-] JSD index is ineffective in detecting phase errors between the predicted and the reference signals, and scarcely able to detect infrequent but large amplitude errors. Instead, it highlights whether the compared time histories assume each value in their range of variation a similar number of times. It is, hence, sensible to errors in the frequency and trend of the predicted signal.
\end{itemize}


\section{Test case description}\label{s:test}
\begin{figure}[ht!]
\centering
\includegraphics[width=0.75\linewidth]{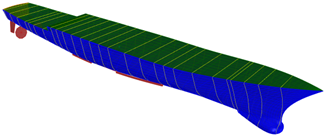}
  \caption{Visualization of the computational grid on the 5415M hull in TEMPEST.}
  \label{fig:grid}
  \bigskip
\end{figure}

The methods presented in this work are applied to the motions of the hull model 5415M, \cref{fig:grid},
a modified form of the destroyer model DTMB 5415 that has been the subject of many model tests and numerical studies, such as the NATO AVT-280 “Evaluation of Prediction Methods for Ship Performance in Heavy Weather” \cite{WalreeVisser2010}.

The present study considers data from free-running simulations of the course keeping of the 5415M in a sea state 7 operating condition, as defined by the World Meteorological Organization, with a nominal significant wave height of 7 m. 
The simulations are conducted in irregular long-crested waves, with a nominal peak period $T_p$ =9.2 s and wave heading of 300 deg, {i.e.}, beam-quartering seas (from 30 degrees aft of the beam, see \cref{fig:scheme}) and a nominal forward speed Froude number Fr = 0.33.
The combination of sea state, heading, and forward speed was selected to achieve a condition close to roll resonance. The resulting hull motions are characterized by significant nonlinearities, establishing a particularly challenging condition to be predicted by DMD-based methods.  

Course-keeping computations are based on the potential flow code TEMPEST \cite{hughes2011tempest}.
TEMPEST was developed by the Naval Surface Warfare Center Carderock Division (NSWCCD) for the simulation of large amplitude ship motions, including capsize \cite{BelknapReed2019}.
TEMPEST uses potential flow modeling to evaluate radiation, diffraction, and incident wave forces on the hull \cite{bandyk2009body}.  
Nonlinear hydrostatic and wave excitation forces are evaluated based on the instantaneous wetted surface.  
Radiation and diffraction forces are evaluated using a body-exact strip-theory method.  
Like many ship motion codes based on potential flow, TEMPEST uses coefficient-based models for forces heavily influenced by viscous effects. Maneuvering forces are modeled using Abkowitz-type stability derivatives \cite{abkowitz1964ship}, and cross-flow lift and drag coefficients that vary with drift angle.  
Input hull maneuvering force terms for model 5415M were evaluated using double-body RANS simulations from NavyFOAM, which is an extension of OpenFOAM for naval hydrodynamics applications \cite{AramField2016,AramKim2017,Kim2017,Aram2018}.  
For propeller forces, TEMPEST applies the widely-used approach of considering thrust coefficients as functions of advance coefficients, which has been implemented using look-up tables with data from \cite{serani2021urans} for model 5415M.  
The influence of the hull on propeller inflow is modeled using straightening coefficients, which were evaluated for the destroyer using double-body RANS simulations.  
Rudder lift forces are evaluated using empirical models from \cite{whicker1958free}, and rudder inflow straightening was assessed using the same double-body RANS simulations used for propellers.
Each irregular seaway was modeled using linear superposition of 100 wave components, that were used to model a JONSWAP spectrum.  
The frequencies of the wave components $\omega_i$ were evenly distributed between 0.41 and 1.47 rad/s, with a constant frequency increment $\Delta \omega_i$ of 0.135 rad/s.  
The amplitude of each wave component was $\zeta_i = \sqrt{2 \: S(\omega_i) \: \Delta \omega_i}$.    
The phase of each wave component was randomly generated from a range between 0 and $2 \pi$.
In the simulations, the free-running ship is kept on course by a simulated proportional derivative control actuating the rudder angle.

\section{Numerical setup}\label{s:nums}
\subsection{Data preprocessing}
Data were collected from 49 independent TEMPEST simulations, each one spanning 6000 time steps over approximately 20 encountered waves.
Twenty-five simulations, roughly half of the data, were reserved as a training set for building the DMD-based ROMs. Twelve simulations are used as a validation set, 
and the remaining 12 simulations are available as a test set. 

For processing ease with DMD, data were downsampled to 32 time steps per nominal wave encounter.
All the analyses are based on normalized data using the Z-score standardization. Specifically, the time histories of each variable are shifted and scaled using the average and standard deviation evaluated on the training set. 

The state vector $\bf x$ used in the DMD analyses is composed of the ship's heave $x_3$, the three rigid body rotation roll $\phi$, pitch $\theta$, and yaw $\psi$, and the surge and sway velocities $v_1$ and $v_2$, respectively. 
In addition, Hankel-DMDc uses the rudder position angle $\alpha$, and the wave elevation measured by a virtual probe placed at the center of gravity of the vessel $\eta_{cg}$ as forcing inputs to the system composing the vector $\bf u$. 

\subsection{System identification}
\begin{figure}[ht!]
    \centering  
        \includegraphics[width=0.5\linewidth]{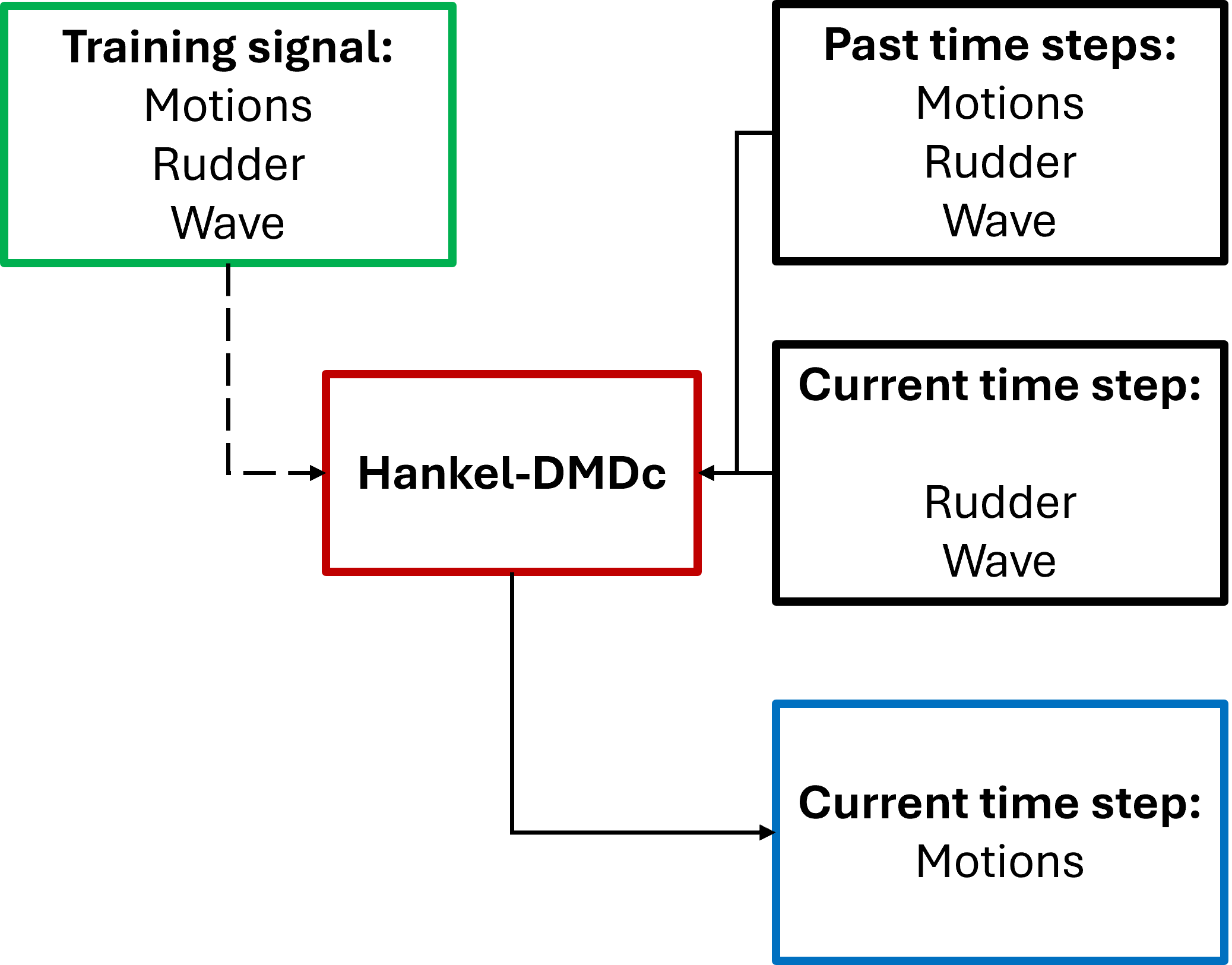}
    \caption{System identification with Hankel-DMDc, conceptual scheme.}
    \label{fig:hdmdc_si1}
\end{figure}

The conceptual scheme for system identification is shown in \cref{fig:hdmdc_si1}
\begin{figure}[ht!]
    \centering   
        \includegraphics[width=\linewidth]{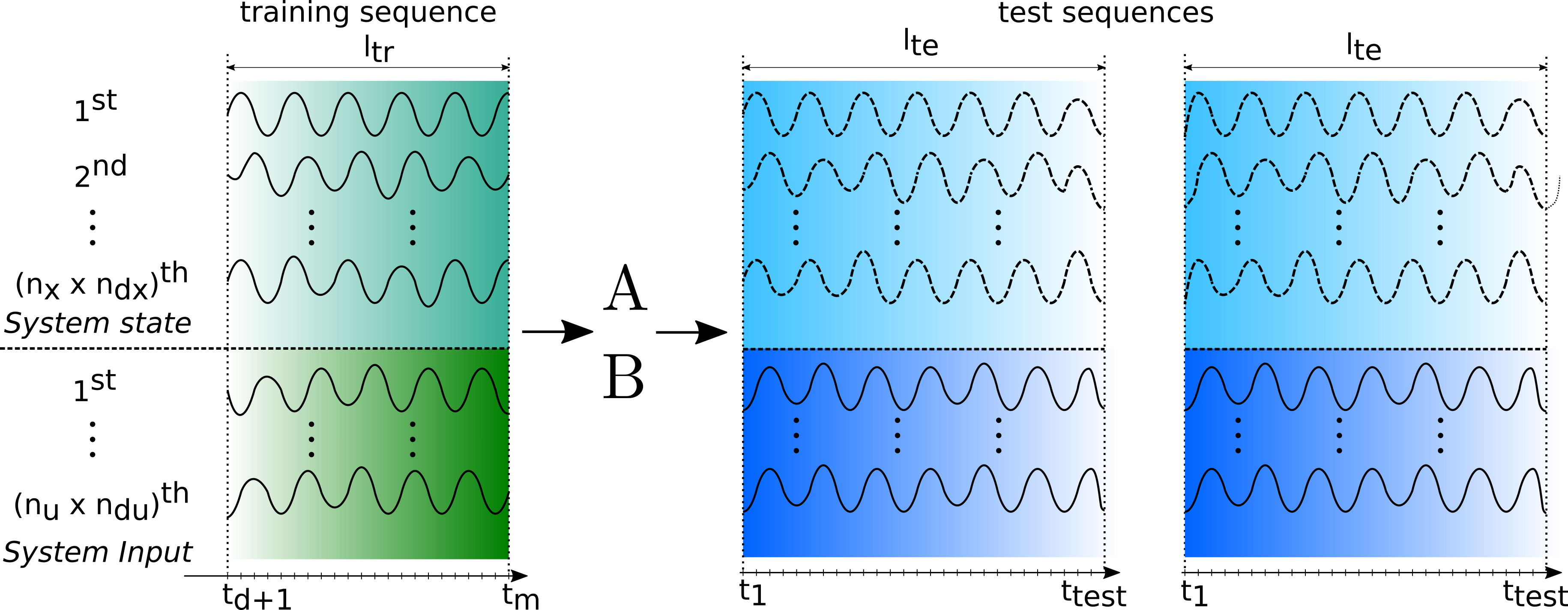} 
    \caption{System identification with Hankel-DMDc, modeling approach sketch.}
    \label{fig:hdmdc_si2}
\end{figure}
The ship's motion is predicted by solving \cref{eq:dmdcdsys} iteratively in time, using matrices $\mathbf{A}$ and $\mathbf{B}$ as identified from the Hankel-DMDc applied to time histories from the training set. 
A linear dynamic system hence approximates the dynamic system describing the ship's motions, forced by the wave elevation and the angular position of the rudder, which values at the current time step are also required, \cref{fig:hdmdc_si2}.

%
The present study considered a full factorial design-of-experiment to assess the effect of the three hyperparameters of Hankel-DMDc on the performance of the resulting models, using six levels for $l_{tr}$ and seven levels for $l_{d_x}$ and $l_{d_u}$. 

The length of a single simulation bounds the sum of the maximum training sequence length and the maximum delay:
\begin{equation}
    l_{tr}^{\text max}+\max(l_{d_x}^{\text max},l_{d_u}^{\text max})\le 20\hat{T},
\end{equation}
where $\hat{T}$ is a reference wave encounter period identified as its average value from the training dataset.
The length of the observation window in validation/test signals $l_{te}$ is also bounded by the maximum delay length and the length of the single simulation:
\begin{equation}\label{eq:ltemax}
    l_{te}^{\text max}+\max(l_{d_x}^{\text max},l_{d_u}^{\text max})\le 20\hat{T}.
\end{equation}

The list of the values explored is reported in \cref{tab:doe_si}, where $n_{tr}$, $n_{d_x}$, and $n_{d_u}$ are the observation and delay lengths, respectively, expressed in signal samples and additional states/inputs number. 
\Cref{eq:ltemax} is used to define the extension of the test signals $l_{te}=15\hat{T}$, allowing to extract one sequence from each simulation in the validation and test sets.
%
\begin{table}[ht!]
    \caption{List of the hyperparameter settings tested for Hankel-DMDc system identification algorithm}\label{tab:doe_si}
    \centering
    \begin{tabular}{lllllllll}
     \toprule
      $l_{tr}$ & [-] & & $1\hat{T}$& $2\hat{T}$& $3\hat{T}$& $5\hat{T}$& $7\hat{T}$& $10\hat{T}$ \\
      $l_{d_x}$  & [-] & $0$ & $0.5\hat{T}$ & $\hat{T}$   & $2\hat{T}$ & $3\hat{T}$ & $4\hat{T}$ & $5\hat{T}$  \\  
      $l_{d_u}$  & [-] & $0$ & $0.5\hat{T}$ & $\hat{T}$   & $2\hat{T}$ & $3\hat{T}$ & $4\hat{T}$ & $5\hat{T}$  \\        
            \midrule
      $n_{tr}$   & [-] &   & 32 & 64 & 96 & 160 & 224 & 320 \\
      $n_{d_x}$  & [-] & 0 & 16 & 32 & 64 & 96  & 128 & 160 \\
      $n_{d_u}$  & [-] & 0 & 16 & 32 & 64 & 96  & 128 & 160 \\
      \bottomrule
    \end{tabular}
\end{table}
%

%
The 294 hyperparameter combinations of the design-of experiment are used to build a different model from each of the 25 simulations in the training set. Each one is tested by evaluating the error metrics defined in \cref{s:metrics} on the 12 validation sequences to assess the results statistically.

\subsection{Bayesian extension}
As highlighted by the authors in previous works \cite{diez2022snh, serani2023, Diez2024, serani2024snh}, the final prediction from DMD-based models may strongly vary for different hyperparameters settings, and no general rule is given for the determination of their optimal values. 
With the aim of including some uncertainty quantification in the prediction, and making the prediction more robust, the Bayesian extension of the Hankel-DMDc considers the hyperparameters as stochastic variables with uniform probability density functions and suitable variation ranges.
The latter are defined after the deterministic analysis, identifying promising combinations based on the statistical results of the three error metrics evaluated on the validation set. Then, 100 Monte Carlo realizations of the set of hyperparameters are considered to determine the expected value of each prediction and its associated standard deviation for the test set.

The same training
time series were used for the deterministic and Bayesian analyses, to ensure their results are comparable. The expected value of the prediction is used to evaluate the error metrics for the Bayesian Hankel-DMDc.

\section{Results and discussion}\label{s:res}
\subsection{Deterministic system identification}\label{s:detres}
\begin{figure*}[ht!]
    \centering
        \includegraphics[width=\linewidth]{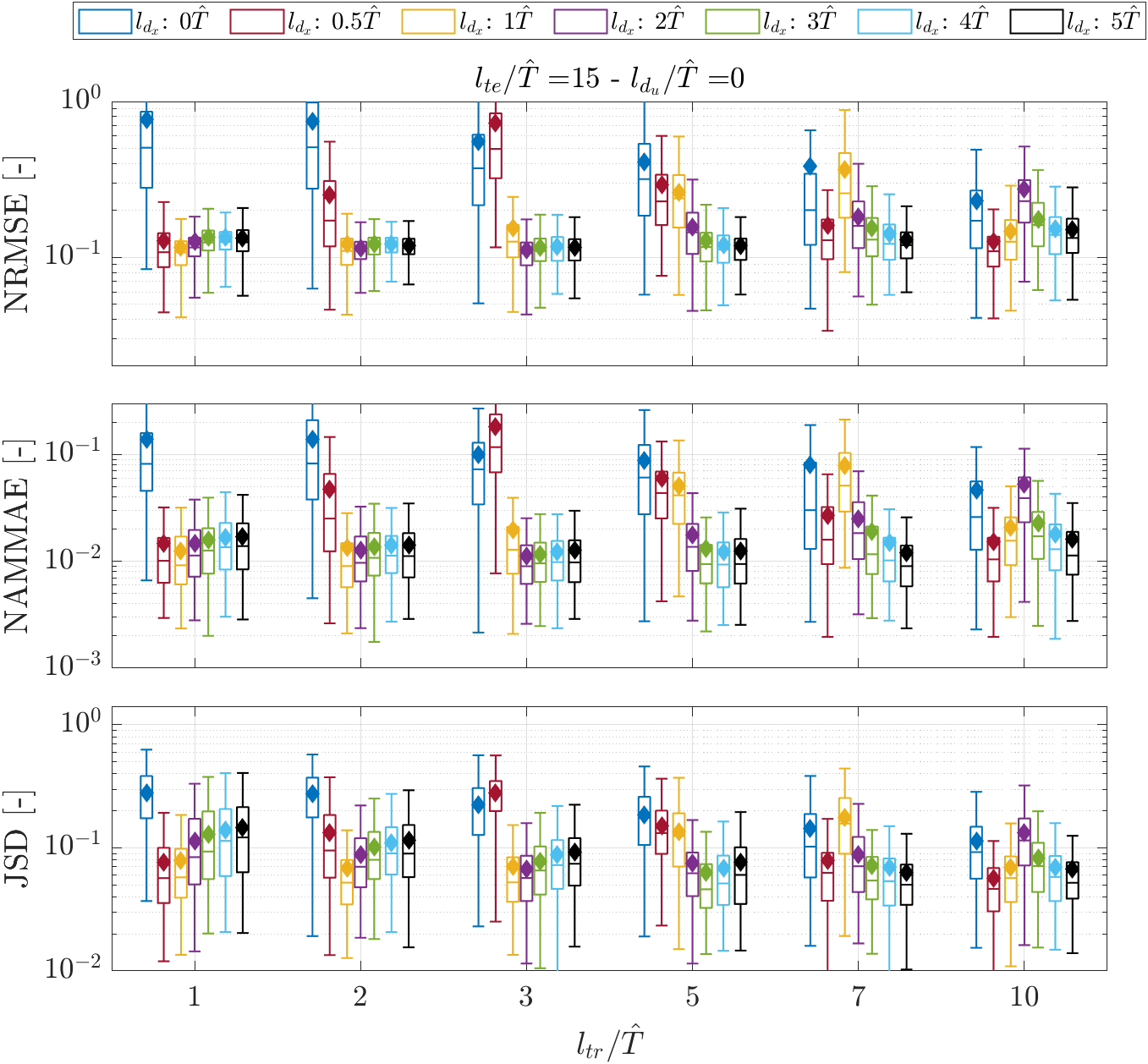}
    \caption{Hankel-DMDc, box plot of error metrics over the validation set for tested $l_{tr}$ and $l_{d_x}$ with $l_{d_u}/\hat{T}=0$. Diamonds indicate the average value of the respective configuration.}
    \label{fig:stat-tempest-1}
\end{figure*}

\begin{figure*}[ht!]
    \centering
        \includegraphics[width=\linewidth]{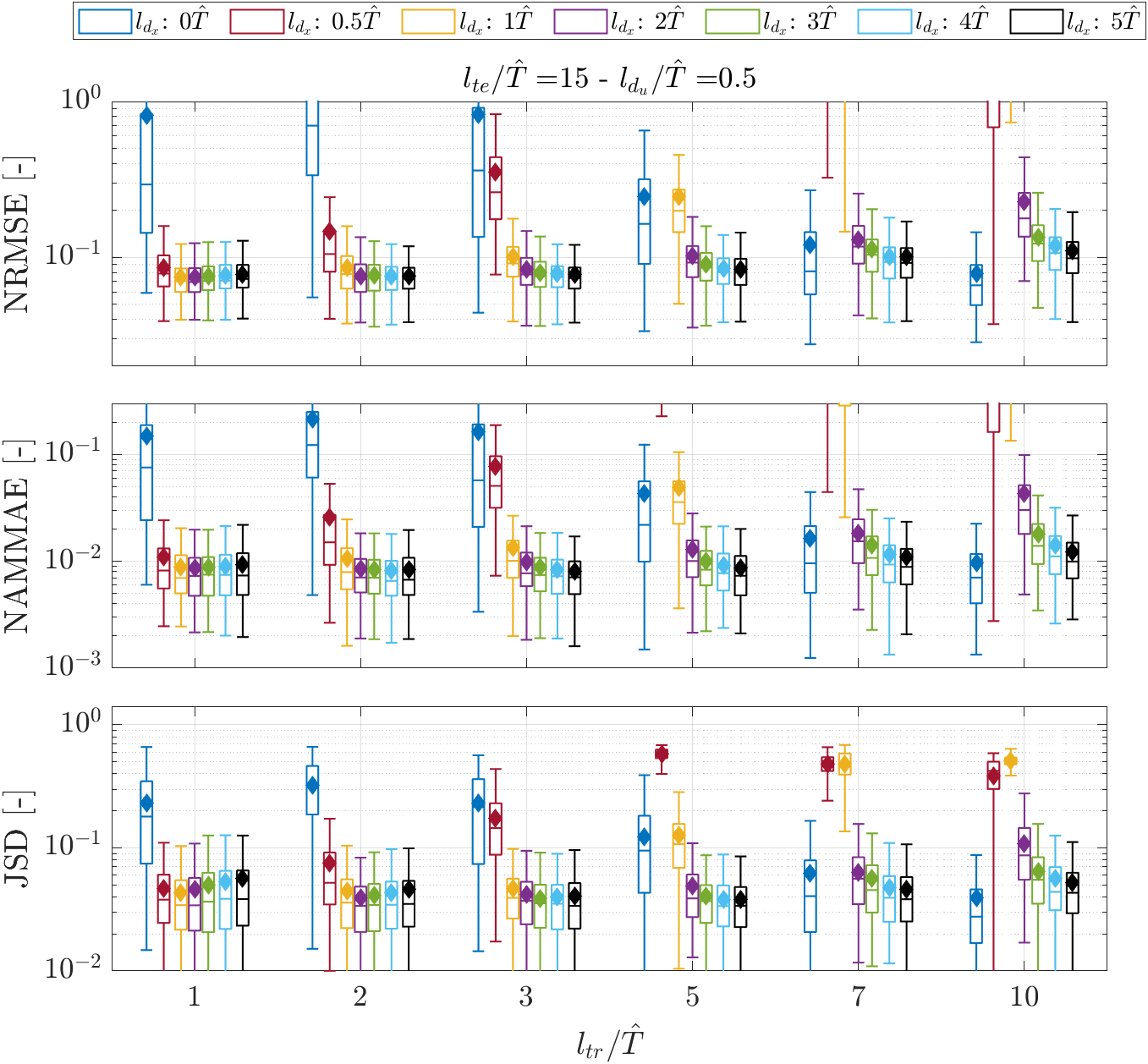}
    \caption{Hankel-DMDc, box plot of error metrics over the validation set for tested $l_{tr}$ and $l_{d_x}$ with $l_{d_u}/\hat{T}=0.5$. Diamonds indicate the average value of the respective configuration.}
    \label{fig:stat-tempest-2}
\end{figure*}

\begin{figure*}[ht!]
    \centering
        \includegraphics[width=\linewidth]{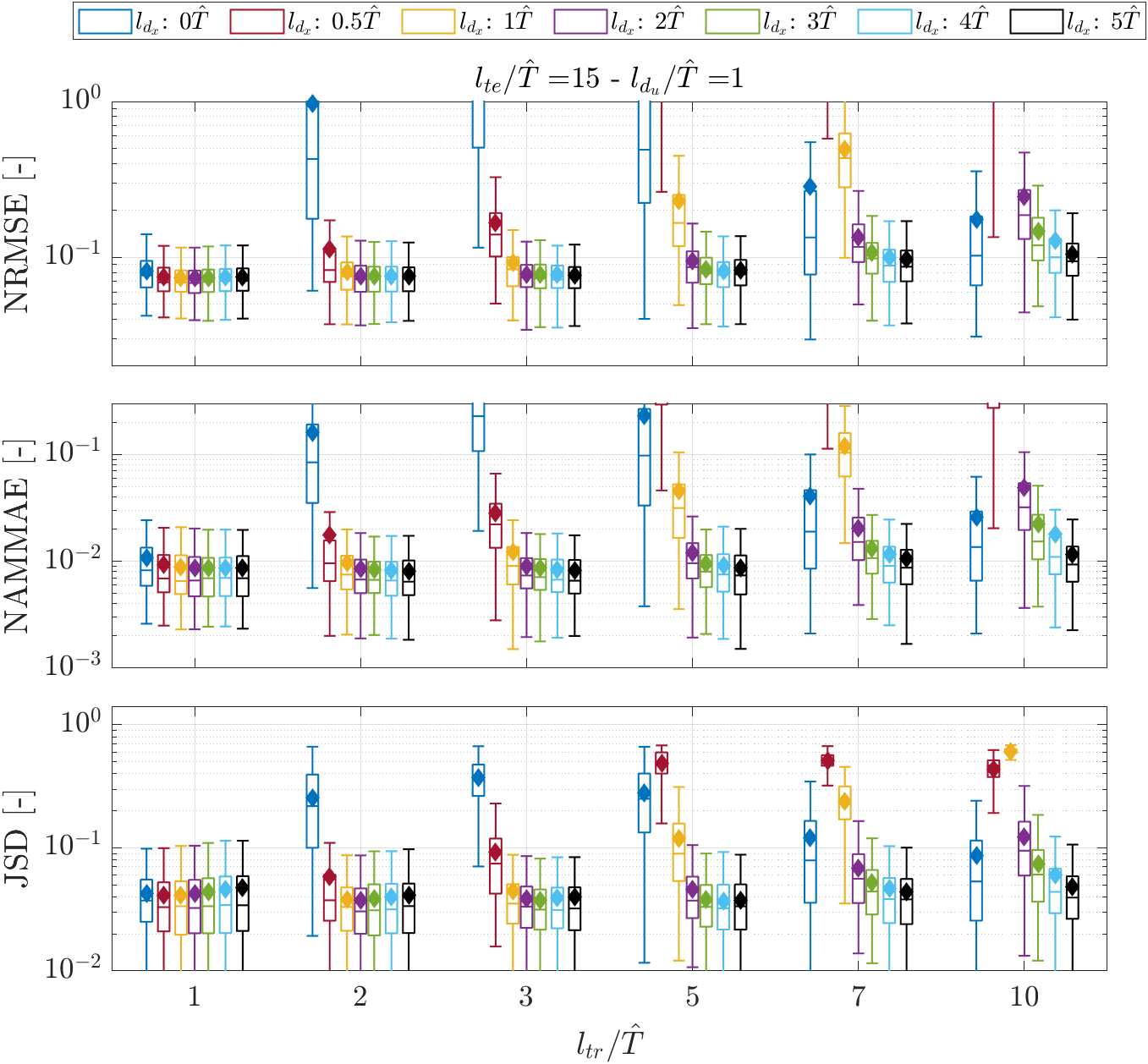}
    \caption{Hankel-DMDc, box plot of error metrics over the validation set for tested $l_{tr}$ and $l_{d_x}$ with $l_{d_u}/\hat{T}=1$. Diamonds indicate the average value of the respective configuration.}
    \label{fig:stat-tempest-3}
\end{figure*}

\begin{figure*}[ht!]
    \centering
        \includegraphics[width=\linewidth]{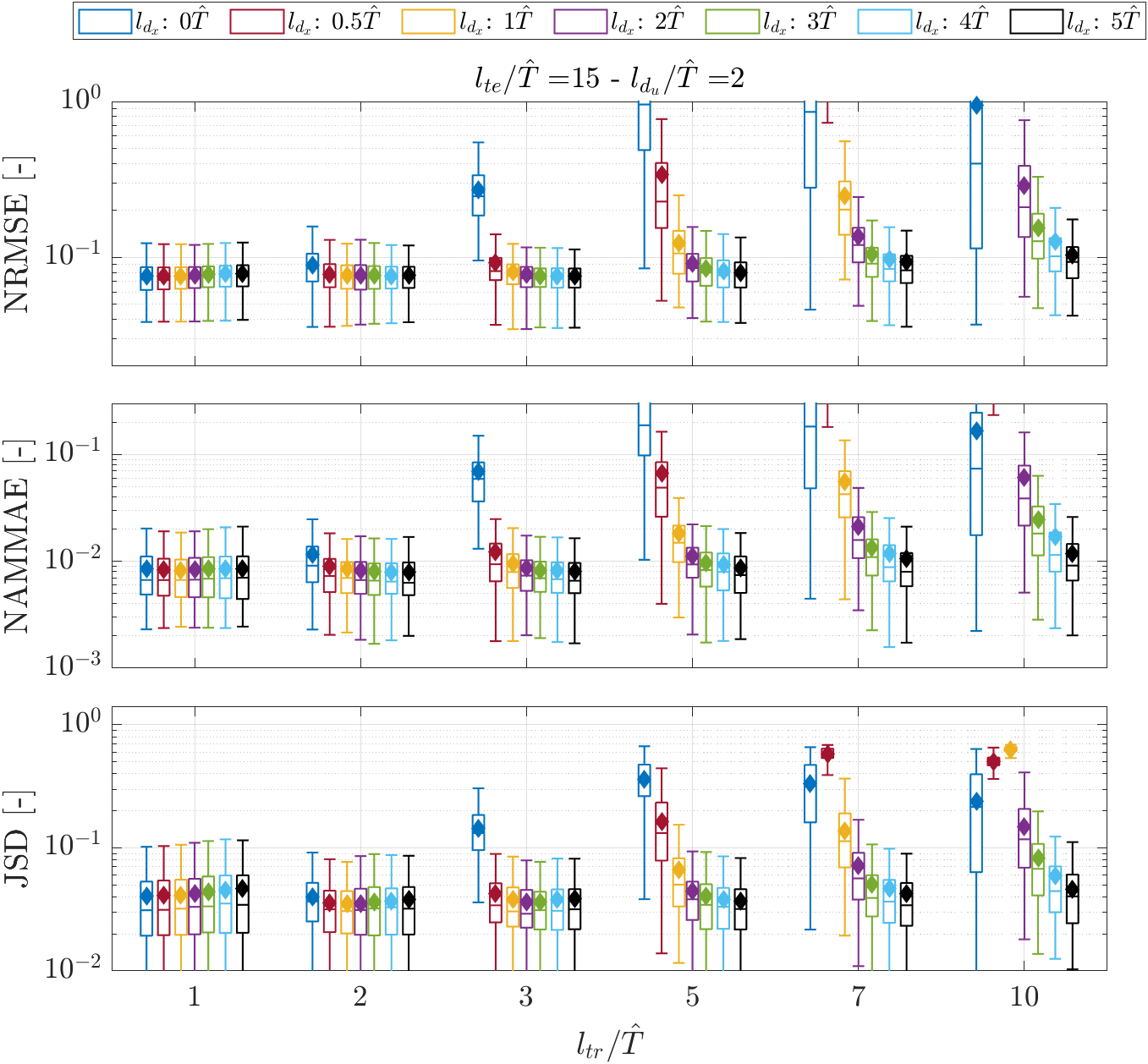}
    \caption{Hankel-DMDc, box plot of error metrics over the validation set for tested $l_{tr}$ and $l_{d_x}$ with $l_{d_u}/\hat{T}=2$. Diamonds indicate the average value of the respective configuration.}
    \label{fig:stat-tempest-4}
\end{figure*}

\begin{figure*}[ht!]
    \centering
        \includegraphics[width=\linewidth]{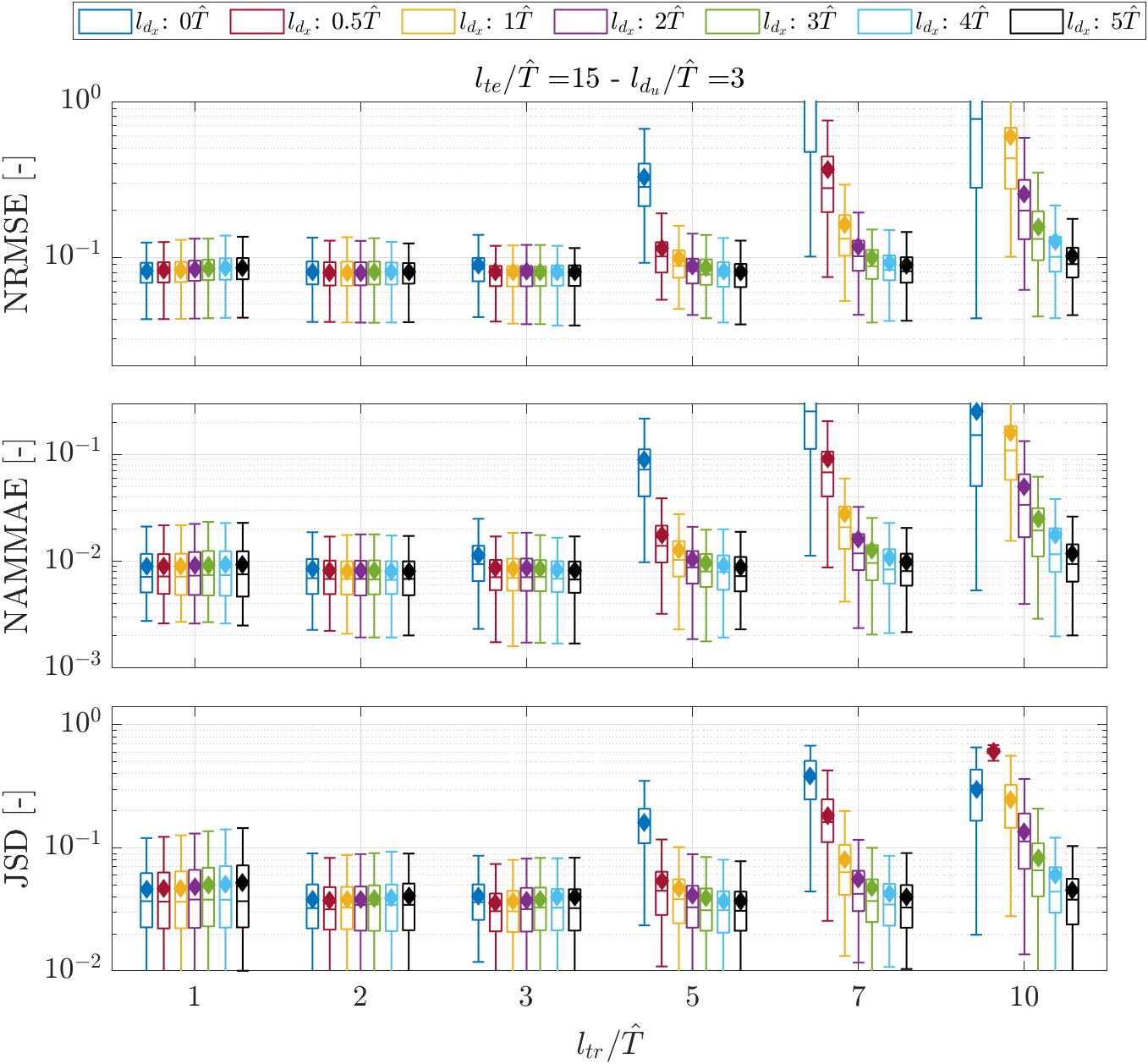}
    \caption{Hankel-DMDc, box plot of error metrics over the validation set for tested $l_{tr}$ and $l_{d_x}$ with $l_{d_u}/\hat{T}=3$. Diamonds indicate the average value of the respective configuration.}
    \label{fig:stat-tempest-5}
\end{figure*}

\begin{figure*}[ht!]
    \centering
        \includegraphics[width=\linewidth]{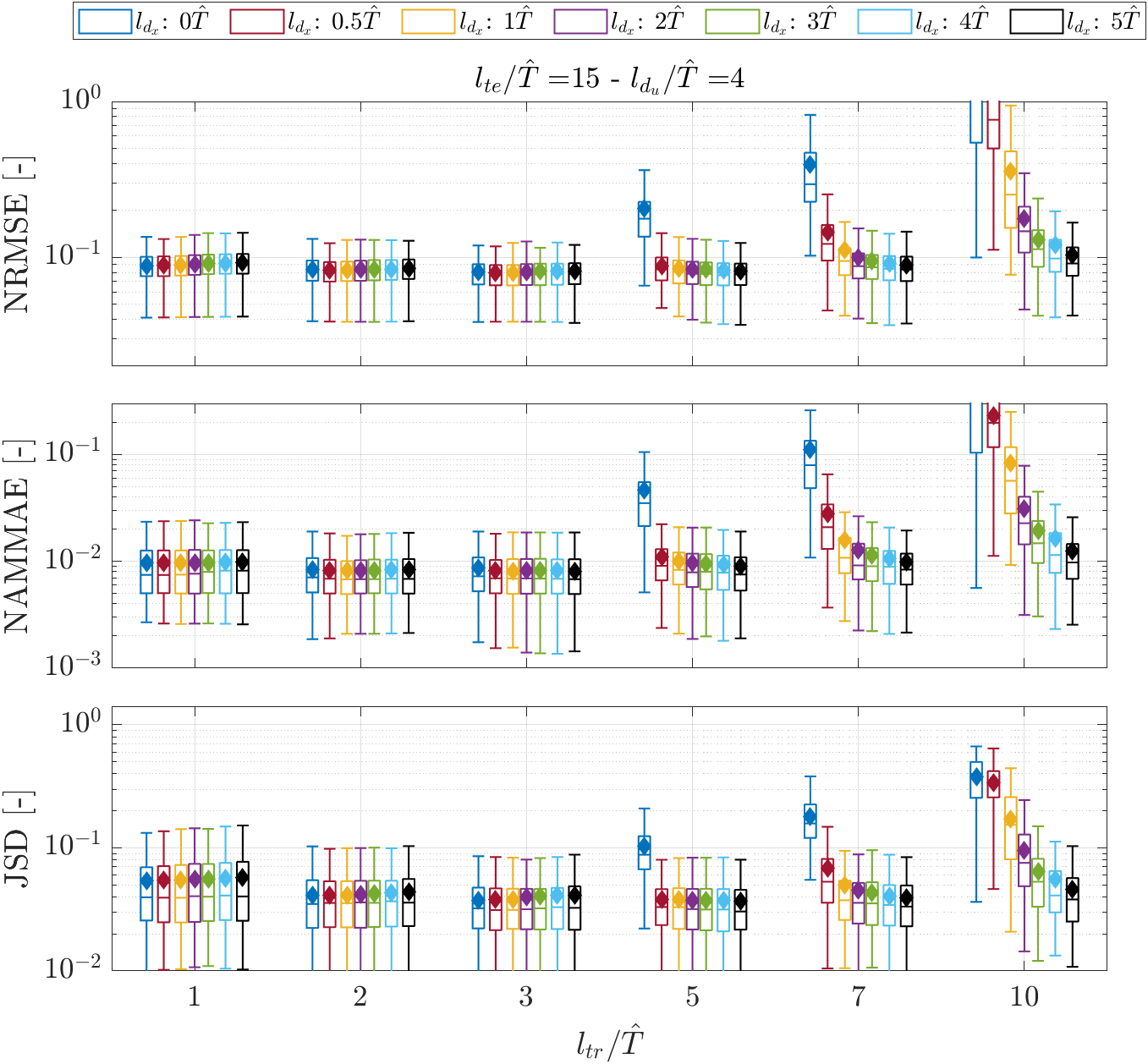}
    \caption{Hankel-DMDc, box plot of error metrics over the validation set for tested $l_{tr}$ and $l_{d_x}$ with $l_{d_u}/\hat{T}=4$. Diamonds indicate the average value of the respective configuration.}
    \label{fig:stat-tempest-6}
\end{figure*}

\begin{figure*}[ht!]
    \centering
        \includegraphics[width=\linewidth]{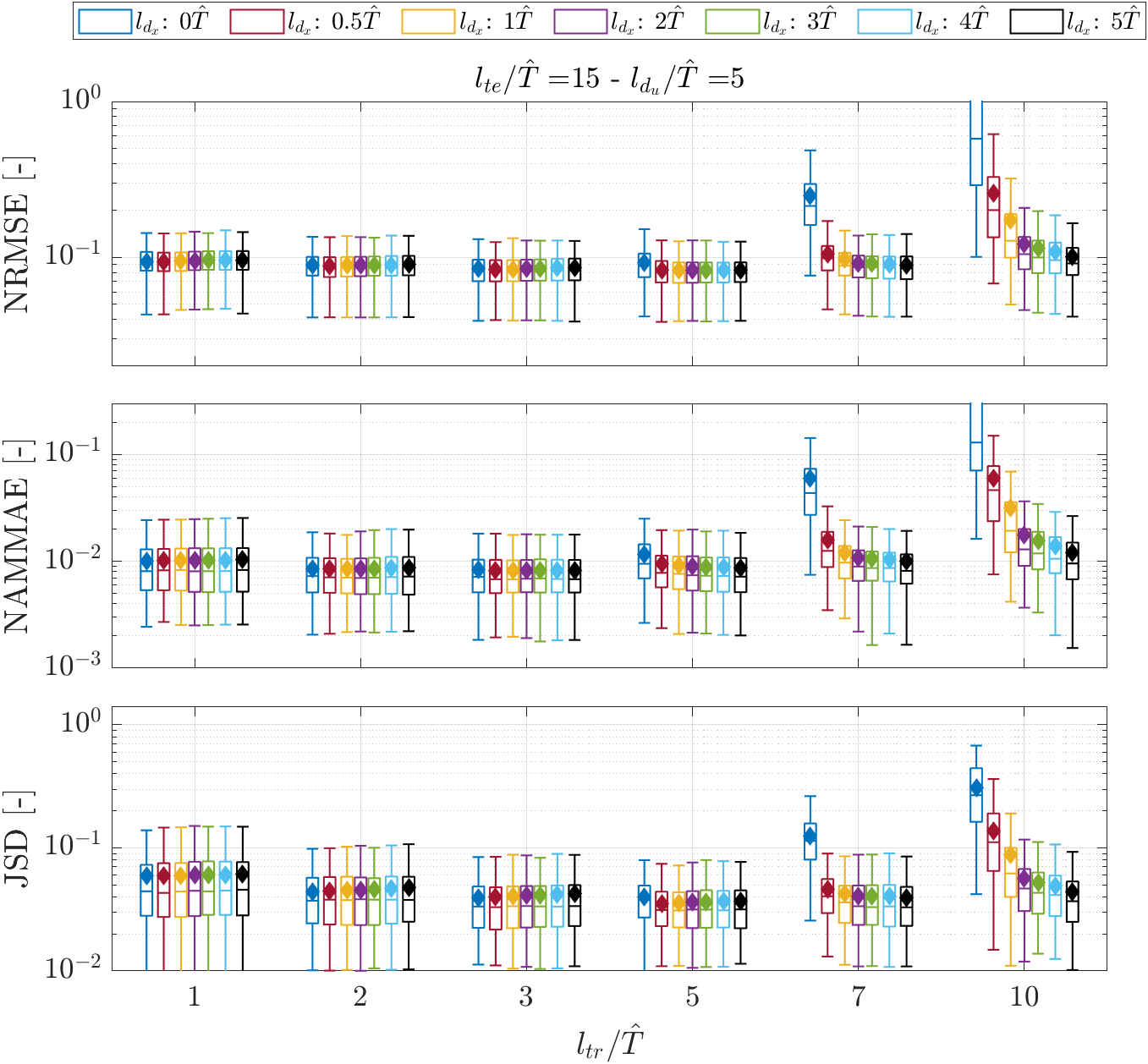}
    \caption{Hankel-DMDc, box plot of error metrics over the validation set for tested $l_{tr}$ and $l_{d_x}$ with $l_{d_u}/\hat{T}=5$. Diamonds indicate the average value of the respective configuration.}
    \label{fig:stat-tempest-7}
\end{figure*}

Results from the deterministic design-of-experiment are reported in terms of box plots graphs in \cref{fig:stat-tempest-1,fig:stat-tempest-2,fig:stat-tempest-3,fig:stat-tempest-4,fig:stat-tempest-5,fig:stat-tempest-6,fig:stat-tempest-7}, combining the outcomes of the 300 training/validation sequences combinations for each hyperparameter configuration. 
The boxes show the first, second (equivalent to the median value), and third quartiles, while the whiskers extend from the box to the farthest data point lying within 1.5 times the interquartile range, defined as the difference between the third and the first quartiles from the box. 
Outliers are not shown to improve the readability of the plot, and a diamond is included for each box indicating the average value of the distribution.
The values of the three error metrics evaluated on the validation set are presented for the prediction length $l_{te} = 15\hat{T}$.


Some configurations of the hyperparameters are characterized by very high values of all the metrics. 
The $\mathbf{A}$ matrices of the pertaining Hankel-DMDc models have been noted to have eigenvalues with positive real parts, causing the predictions to be unstable. 
As pointed out by \cite{rains2024}, DMDc algorithms are particularly susceptible to the choice of the model dimensions and prone to identify spurious unstable eigenvalues. 
With DMD and Hankel-DMD, this phenomenon can be effectively mitigated by projecting the unstable discrete-time eigenvalues onto the unit circle. However, with DMDc and Hankel-DMDc, the $\mathbf{B}$ matrix is also affected by the identification of unstable eigenvalues, and no simple stabilization procedure is available.  

\Cref{tab:detbestconf15} reports the value of the hyperparameters for the configurations producing the best average value of NRMSE, NAMME, and JSD. 
Those lay in a subdomain of variation for the hyperparameters defined as $1\hat{T}\le l_{tr} \le 3\hat{T}$, $1\hat{T}\le l_{d_x} \le 5\hat{T}$, and $1\hat{T}\le l_{d_u} \le 5\hat{T}$.
In general, it is observed from \cref{fig:stat-tempest-1,fig:stat-tempest-2,fig:stat-tempest-3,fig:stat-tempest-4,fig:stat-tempest-5,fig:stat-tempest-6,fig:stat-tempest-7} that the configurations included in this subdomain are characterized by low values for all three error metrics.
\begin{table}
\caption{Resume of the best hyperparameters configurations as identified by the deterministic design of experiment, $l_{te}=15 \hat{T}$, and Bayesian setup.}\label{tab:detbestconf15}
\begin{tabular}{lllllll}
                                   & NRMSE & NAMMAE & JSD   & $l_{tr}$       & $l_{d_x}$      & $l_{d_u}$     \\
                                   & (avg) &  (avg) & (avg) &                &                &               \\
      \midrule
       $\text{Best}_\text{NRMSE}$  &0.0725 &0.00837 &0.0466 & 1$\hat{T}$     & 2$\hat{T}$     & 1$\hat{T}$    \\
       $\text{Best}_\text{NAMMAE}$ &0.0744 &0.00768 &0.0425 & 2$\hat{T}$     & 5$\hat{T}$     & 2$\hat{T}$    \\
       $\text{Best}_\text{JSD}$    &0.0753 &0.00798 &0.0374 & 2$\hat{T}$     & 1$\hat{T}$     & 2$\hat{T}$    \\
       Bayesian                    &0.0692 &0.00734 &0.0393 & [1-3]$\hat{T}$ & [1-5]$\hat{T}$ & [1-2]$\hat{T}$\\
     \midrule
\end{tabular}
\end{table}

This contrasts the results of previous works by the authors \cite{serani2024snh}, where it was noted that higher values of $l_{tr}$ 
(around $200\hat{T}$) were required for the system identification algorithm to reach satisfactory accuracy. 
The discrepancy may be ascribed to the introduction in this work of delayed copies of the state and input in the data matrices, \textit{i.e.}, the full exploitation of the Hankel extension to the DMDc algorithm. In fact, limiting the analysis to hyperparameters configurations with $l_{d_x},l_{d_u}=0$, \textit{i.e.}, the standard DMDc as in the mentioned reference, the errors are notably higher, and decreasing for longer training, compatibly with \cite{serani2024snh}.

\begin{figure*}[ht!]
    \centering
        \includegraphics[width=\linewidth]{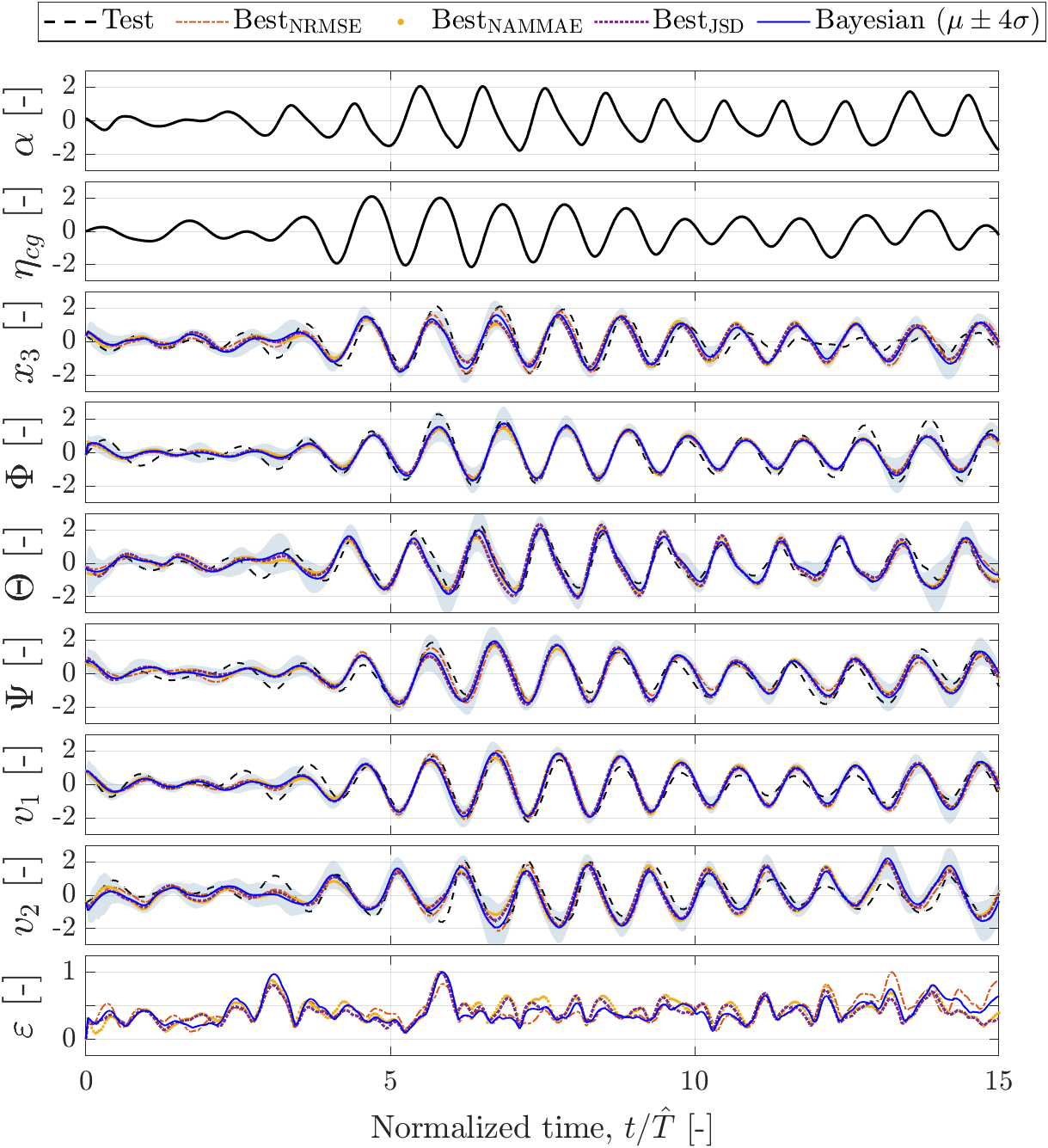}
        \caption{Standardized time series prediction by deterministic (hyperparameters for best average metrics) and Bayesian Hankel-DMDc. Selected sequence 1.} \label{fig:fore1-tempest}    
\end{figure*}
\begin{figure*}[ht!]
    \centering
        \includegraphics[width=\linewidth]{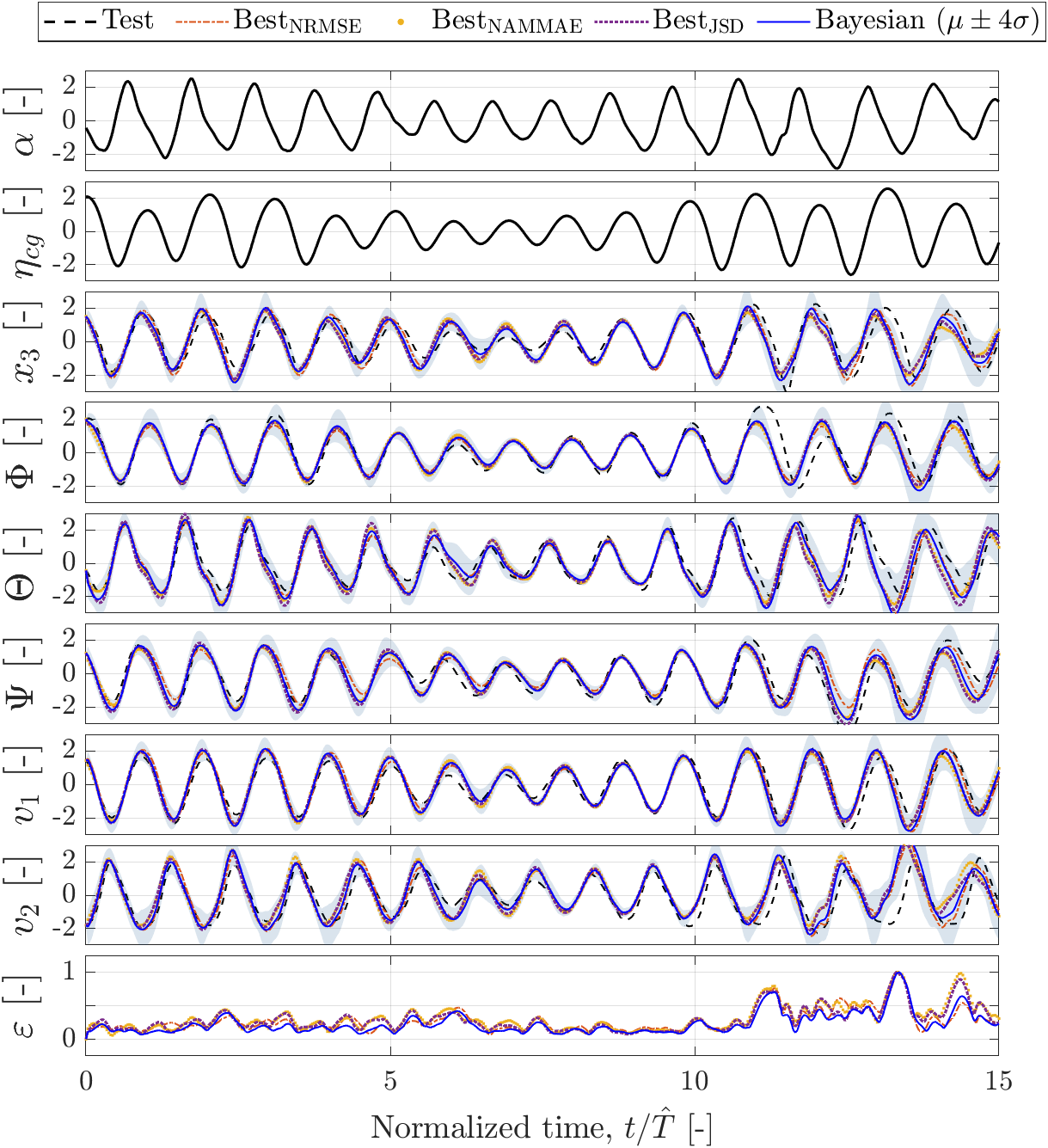}
        \caption{Standardized time series prediction by deterministic (hyperparameters for best average metrics) and Bayesian Hankel-DMDc. Selected sequence 2.} \label{fig:fore2-tempest}   
\end{figure*}
\begin{figure*}[ht!]
    \centering
        \includegraphics[width=\linewidth]{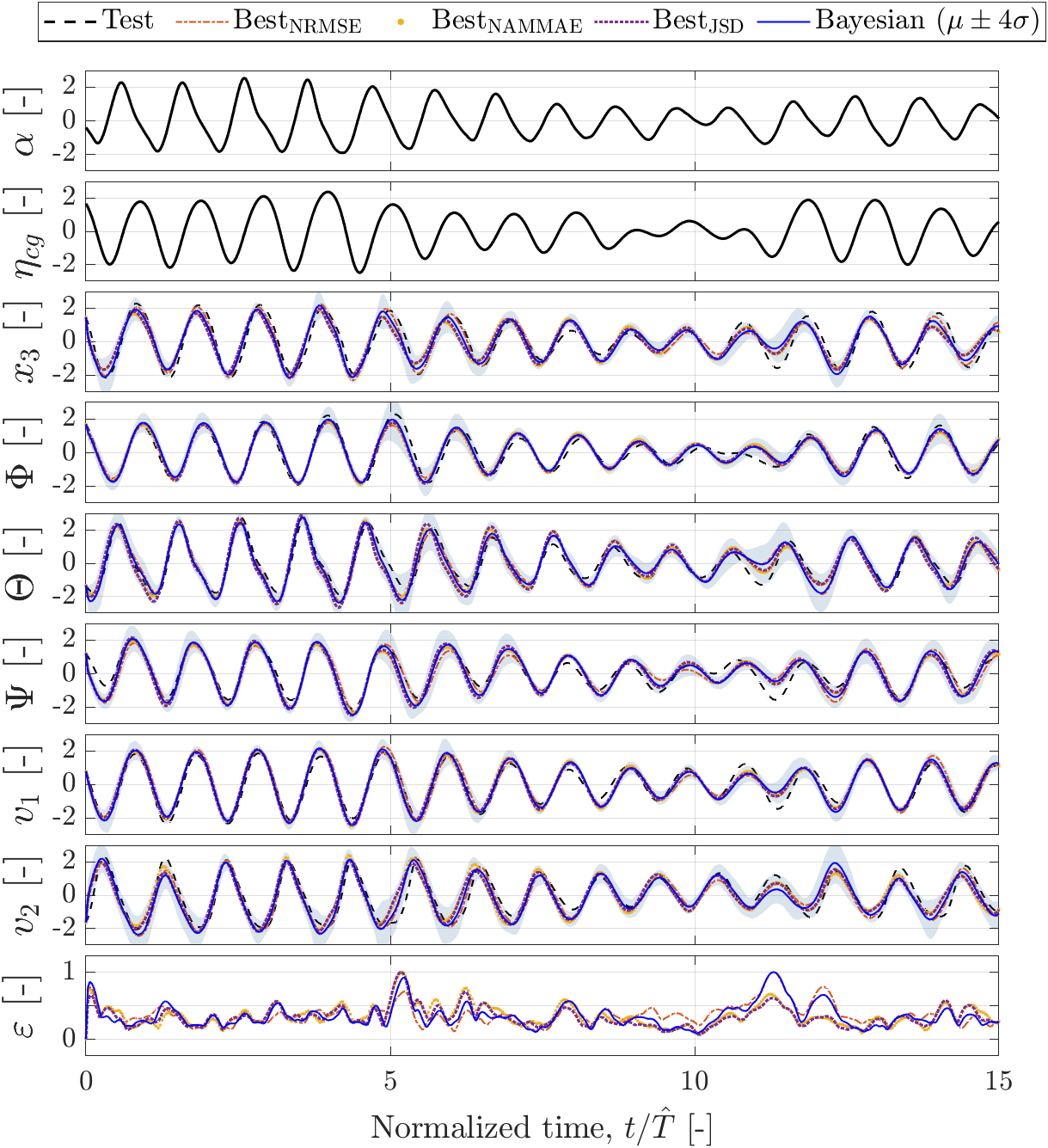}
        \caption{Standardized time series prediction by deterministic (hyperparameters for best average metrics) and Bayesian Hankel-DMDc. Selected sequence 3.} \label{fig:fore3-tempest}    
\end{figure*}
\begin{figure*}[ht!]
    \centering        
       \includegraphics[width=\linewidth]{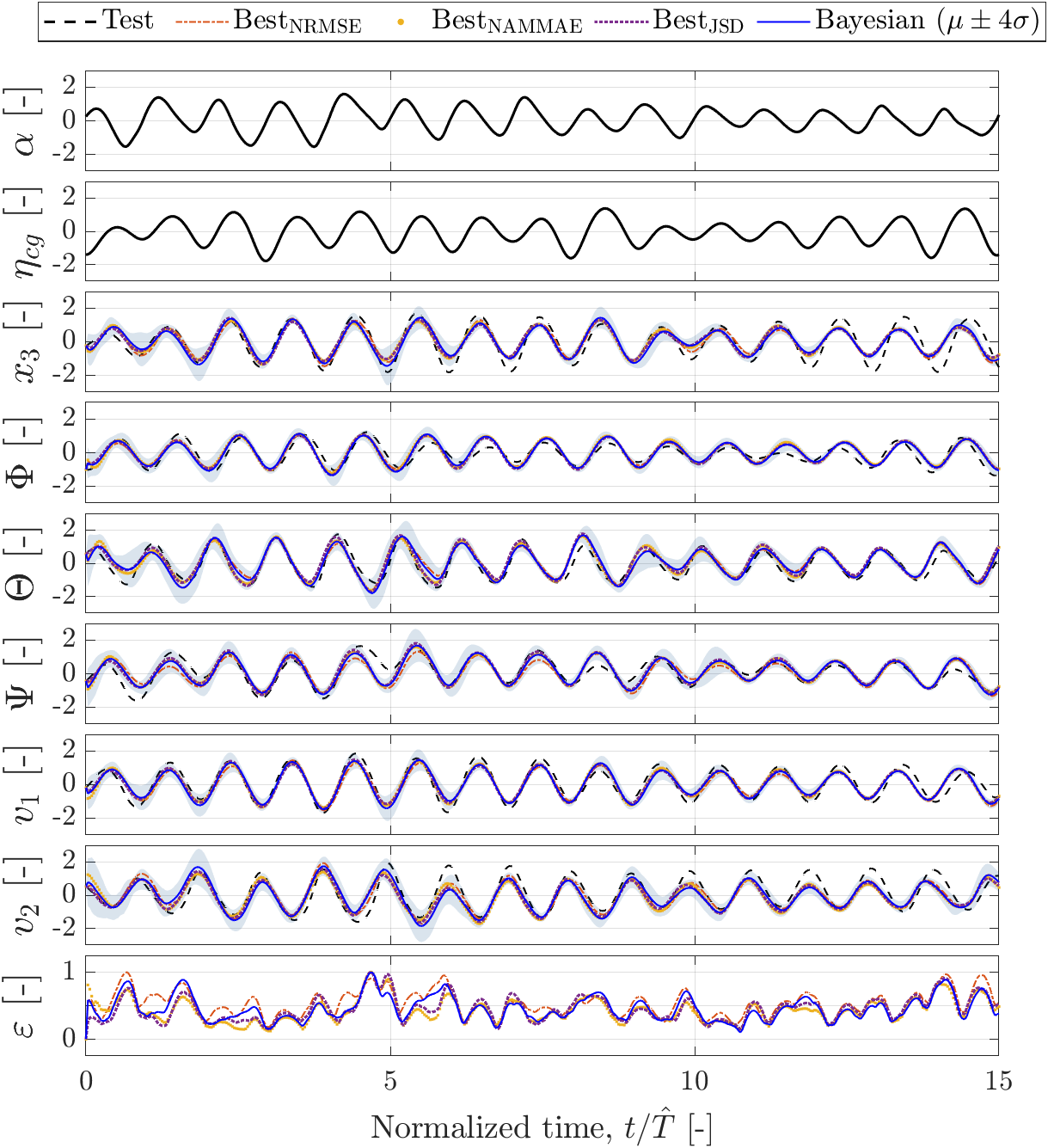}
        \caption{Standardized time series prediction by deterministic (hyperparameters for best average metrics) and Bayesian Hankel-DMDc. Selected sequence 4.} \label{fig:fore4-tempest}           
\end{figure*}

\Cref{fig:fore1-tempest,fig:fore2-tempest,fig:fore3-tempest,fig:fore4-tempest} show the comparison for some test sequences between reference signals (black dashed lines) and the deterministic Hankel-DMDc predictions using the hyperparameters in \cref{tab:detbestconf15}, \textit{i.e.}, providing the best average value of NRMSE (dash-dotted orange line), NAMMAE (fine dotted yellow line), and JSD (dotted purple line).
In addition to the inputs and predicted states, the figures show the time evolution of the squared root difference between the reference and predicted signal, averaged on the variables and normalized by its maximum value:
\begin{equation}
    \varepsilon = \frac{\bar{\varepsilon}(t)}{\max(\bar{\varepsilon}(t))}, \text{ with:}\quad
    \bar{\varepsilon}(t) = \frac{1}{N} \sum_i^N \sqrt{\left(\tilde{x_i}(t)-x_i(t)\right)^2}
\end{equation}
A qualitative analysis shows good agreement between the prediction and the ground truth for the entire observation window. No degradation trend in accuracy is noted while advancing with the prediction time.

The reported examples contain time histories in which amplitudes strongly vary in the observation window, while others show almost stable oscillations. 
The Hankel-DMDc models are able to follow the variations in all the situations, from small to large amplitude and vice versa. This shows the fundamental role of the forcing term in the solution and confirms the effectiveness of the DMDc extension for the long-term prediction task.

\subsection{Bayesian system identification}\label{s:bayres}
The Bayesian extension of the Hankel-DMDc algorithm is obtained by exploiting the insights on the hyperparameters derived from the deterministic analysis.
In \cref{s:detres}, the hyperparameters subdomain $1 \le l_{tr}/\hat{T} \le 3$, $1 \le l_{d_x}/\hat{T} \le 5$, and $1\le l_{d_u}/\hat{T} \le 2$ is identified as promising for containing the configurations obtaining the best average values of the metrics, and showing overall good prediction accuracies. 
For these reasons, it is considered here for the Bayesian analysis.
The three hyperparameters are, hence, treated as probabilistic variables, uniformly distributed in their ranges, $l_{tr}/{\hat{T}}$~$\sim$~$\mathcal{U}(1,3)$, $l_{d_x}/{\hat{T}}$~$\sim$~$\mathcal{U}(1,5)$, and $l_{d_u}/{\hat{T}}$~$\sim$~$\mathcal{U}(1,2)$. It shall be noted that the actual $n_{d_x}$ and $n_{d_u}$ are taken as the nearest integers from the calculated values.

\Cref{fig:fore1-tempest,fig:fore2-tempest,fig:fore3-tempest,fig:fore4-tempest} compare Bayesian predictions with the test signal and the best deterministic predictions. The blue solid line represents the expected value of the Bayesian prediction, and Chebyshev’s inequality is used for the shaded area representing uncertainty, with a coverage factor equal to 4 (93.75\% confidence interval). 

A qualitative analysis indicates that the expected values of the Bayesian predictions reach a good level of accuracy, typically improving or matching the deterministic predictions. 
The ground truth almost always falls inside the uncertainty band of the prediction. 
As observed for the deterministic analysis, i) no accuracy degradation trend with the prediction time is noted, and ii) the Bayesian Hankel-DMDc is effective in several scenarios, including stable oscillations and growing/decaying signal portions combined in various orders.

\begin{figure*}[ht!]
    \centering
    \includegraphics[width=0.8\linewidth]{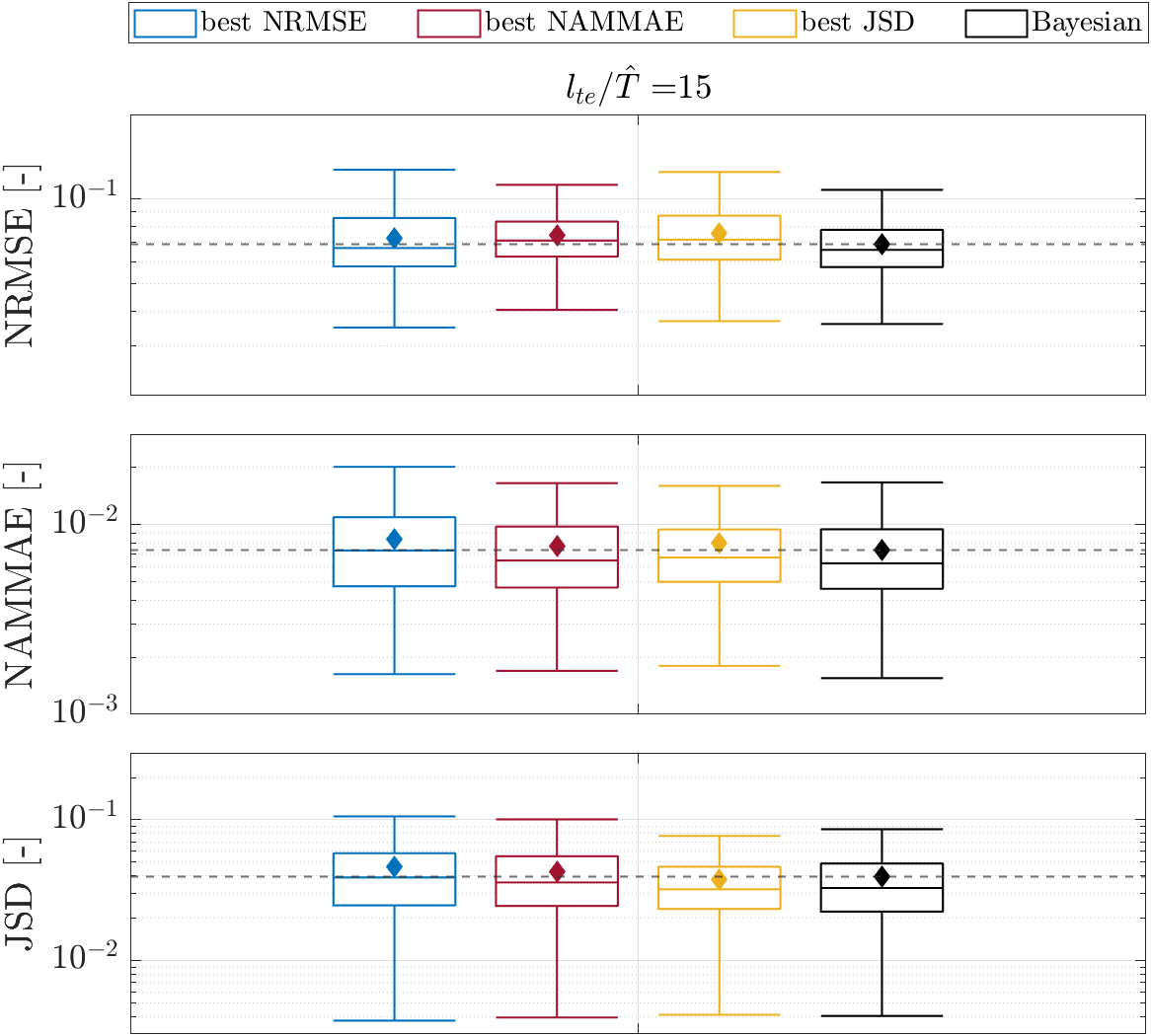}
    \caption{Hankel-DMDc deterministic best$_\text{NRMSE}$, best$_\text{NAMMAE}$, and best$_\text{JSD}$ configurations vs Bayesian Hankel DMDc, box plots of error metrics. Diamonds indicate the average value of distributions in the box plots, and horizontal dashed lines indicate the average value of the metrics for the Bayesian expected prediction.}
    \label{fig:bstat-tempest}
\end{figure*}

A quantitative comparison between the deterministic and Bayesian approaches is obtained by evaluating the statistics of the error metrics for the expected value of the Bayesian prediction over the 300 training/validation combinations and comparing it with the deterministic best average NRMSE, NAMMAE, and JSD solutions. 
\Cref{fig:bstat-tempest} presents the results as box plots. 
The dashed horizontal line in each plot indicates the average value of the error metric obtained by the Bayesian algorithm, also reported in \cref{tab:detbestconf15}. 
The Bayesian expected solution obtains the best average NRMSE and NAMME values, and the second-best JSD value, combining the qualities shown by the different best deterministic setups.
%
%

\begin{figure*}[ht!]
    \centering        
       \includegraphics[width=0.7\linewidth]{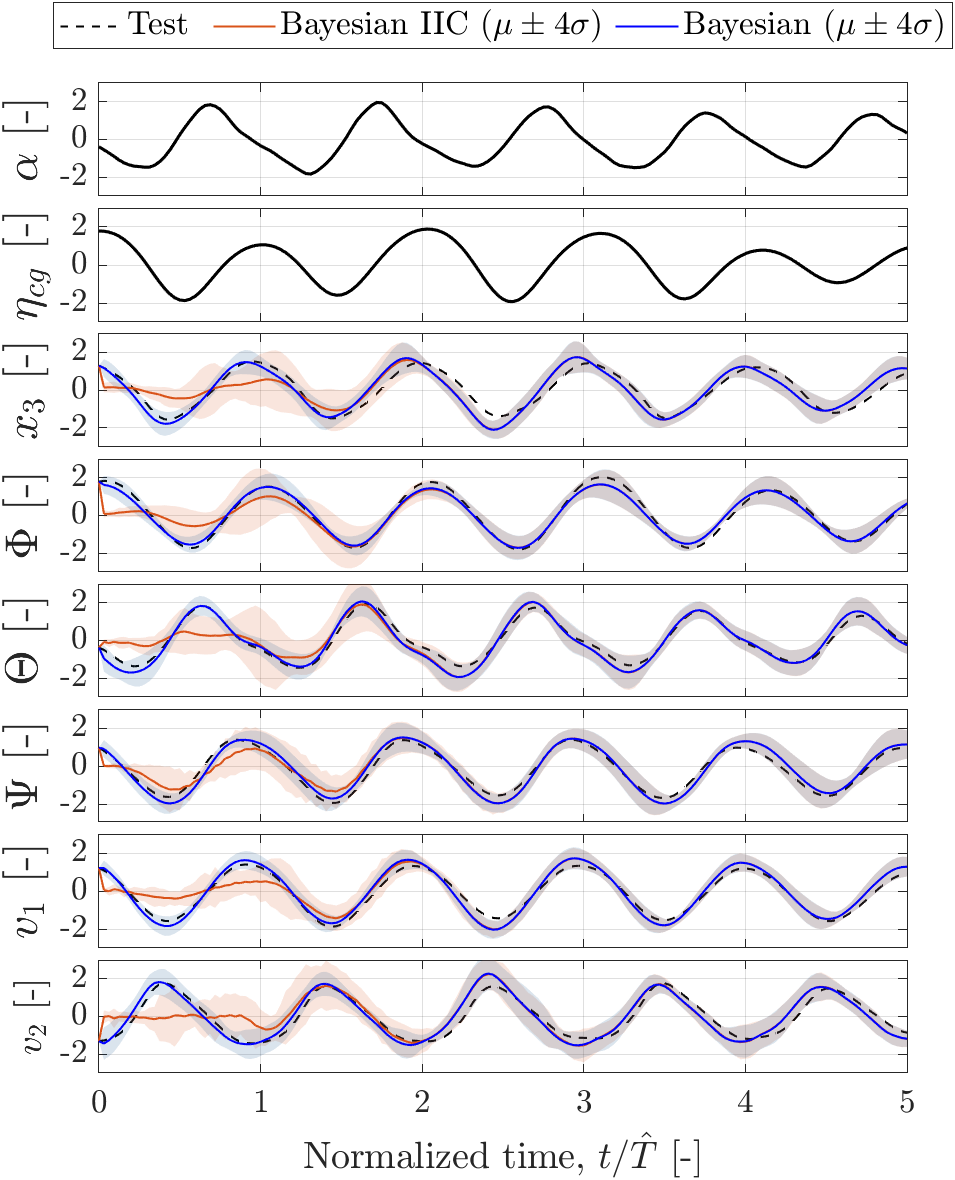}
        \caption{Bayesian Hankel-DMDc, comparison between evolution from complete and incomplete initial conditions (IIC).} \label{fig:cii-tempest}
\end{figure*}

%
The ROMs obtained with (Bayesian) Hankel-DMDc can be used as surrogate models replacing the data source used for their generation (\textit{e.g.}, CFD simulations). In this case, it shall be considered that no information about the past state is available at the first time step, \textit{i.e.}, no delayed state is known. In addition, the same holds for the delayed inputs: the vectors $\mathbf{u}_{j-k}$, $k=1,z$ composing $\hat{\mathbf{u}}_j$ become available only after $j\ge k$ time steps. 
A possible approach is to replace the unknown information with zeros and let the system evolve from an incomplete initial condition (IIC). 
As shown in \cref{fig:cii-tempest}, the predictions obtained from the incomplete and complete initial conditions rapidly become indistinguishable.
Reflecting the limited memory of dynamical systems, the influence of the initial conditions fades within approximately the first three wave encounter periods. Hence, as a rule-of-thumb, one can discard a transient $5\hat{T}$ long, the maximum between the upper bounds of $l_{d_x}$ and $l_{d_u}$. 

Considering this application, the DMD-based ROMs are remarkably efficient from a computational point of view: the calculation of the $\mathbf{A}$ and $\mathbf{B}$ matrices requires less than a second on a laptop mounting an i5-1235U CPU, and 16 GB RAM ($\mu=0.6878$, $\sigma=0.0857$ seconds), and the evaluation of a 15$\hat{T}$ long test time sequence is performed in hundredths of a second ($\mu=0.0438$, $\sigma=0.0045$ seconds). The training set can be as small as a single sequence 5 wave encounter periods long, enabling accurate prediction of arbitrary new sequences in the same operational conditions.
%

\begin{figure*}
    \centering
    \includegraphics[width=0.7\linewidth]{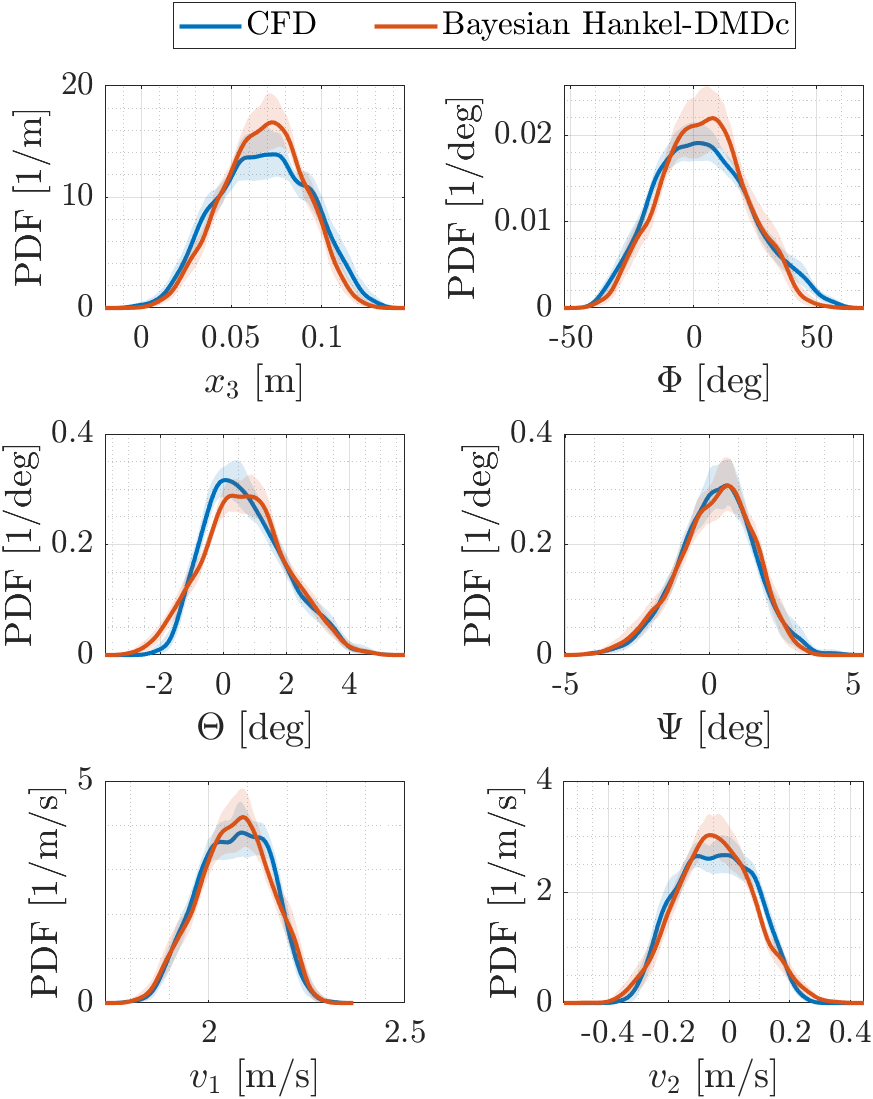}
    \caption{Probability density function comparison between CFD data and Bayesian Hankel-DMDc prediction expected value. Shaded areas indicate the 95\% confidence interval of the two PDFs.}\label{fig:PDFallvars}
\end{figure*}

The capability of the Bayesian Hankel-DMDc ROM to be used as an alternative to the CFD solver is assessed by statistically comparing the probability density functions (PDF) of the time series arising from the two methods. 
A moving block bootstrap (MBB) method is applied to define the time series to be analyzed, following \cite{serani2021urans}, introducing uncertainty in the PDFs estimation using 100 bootstrapped series. 
The PDF of each time series is evaluated from the expected value of the prediction obtained by the ROM and the original simulations using kernel density estimation \cite{Miecznikowski2010} as:
\begin{equation}
    \text{PDF}\left(x,y\right) = \frac{1}{\mathcal{T} h}\sum_{i=1}^{\mathcal{T}} K \left(\frac{y -x_i}{h}\right).
\end{equation}
In particular, $K$ is a normal kernel function defined as
\begin{equation}
    K\left(\xi\right) = \frac{1}{\sqrt{2 \pi}} \exp{\left(-\frac{\xi^2}{2}\right)},
\end{equation}
where $h=\sigma(x)\mathcal{T}^{-1/5}$ is a bandwidth \cite{Silverman2018}.
Quantile function $q$ is evaluated at probabilities $p=0.025$ and $0.975$ defining the lower and upper bounds of the 95\% confidence interval as $U_{PDF(\xi,y)} = PDF(\xi,y)_{q=0.95} - PDF(\xi,y)_{q=0.025}$.
\Cref{fig:PDFallvars} shows the expected value of the PDFs over the bootstrapped series as solid lines and the 95\% confidence interval as shaded areas.
Results show a satisfactory agreement between the distributions obtained from ROM and CFD. 
The confidence intervals of the Bayesian Hankel-DMDc predictions adequately cover the PDF of the CFD simulations, indicating that the predictions are accurate and reliable.
The statistics of the JSD evaluated on the bootstrapped time series for the six variables are reported in \cref{tab:mbbjsd} to quantify the similarity between the CFD and DMD distributions in \cref{fig:PDFallvars}. The analysis confirms the closeness between the Bayesian Hankel-DMDc prediction and the CFD simulations. The JSD expected value is small and close to the lower bound of the interval confidence, confirming the accuracy and reliability of the ROM predictions.
\begin{table}[ht!]
    \centering
    \caption{Expected value and 95\% confidence lower bound, upper bound, and interval of JSD of bootstrapped time series}
    \label{tab:mbbjsd}  
    \begin{tabular}{lllll}
             & JSD($\xi$) & \\
    $\xi$    & EV  & q=0.025& q=0.975 & U \\
             \midrule
    $x_3$    & 0.0088 & 0.0075 & 0.0234 & 0.0159\\
    $\Phi$   & 0.0090 & 0.0079 & 0.0211 & 0.0132\\
    $\Theta$ & 0.0103 & 0.0095 & 0.0191 & 0.0096\\
    $\Psi$   & 0.0055 & 0.0048 & 0.0131 & 0.0083\\
    $v_1$    & 0.0045 & 0.0038 & 0.0118 & 0.0080\\
    $v_2$    & 0.0082 & 0.0078 & 0.0149 & 0.0071\\
    \midrule
    avg      & 0.0077 & 0.0068 & 0.0172 & 0.0103\\
    \midrule
    \end{tabular}
\end{table}
%

\section{Conclusion}\label{s:concl}
This study applied the Hankel-DMDc and its novel Bayesian extension for the model-free (data-driven and equation-free) system identification and prediction of ship motions for a vessel operating in irregular waves. Results have been obtained for a complex sea state, characterized by significant nonlinear interactions between waves and hull motions. 

The inclusion of delayed copies of the state in the Hankel augmented DMD is effective in capturing the important nonlinear features of the dynamic system.
Comparing the performances of the deterministic and Bayesian approaches, the analyses demonstrated the effectiveness of the proposed methodology using data from numerical simulations, reaching remarkable predicting accuracy up to 15 wave encounter periods.
The absence of a clear trend in the evolution of the prediction error inside the test window suggests that the method successfully identifies the analyzed dynamic system, obtaining a ROM that can keep the same accuracy level for an even longer observation time.

The deterministic and Bayesian equation-free dynamic models are obtained with small data and a direct procedure from the training set.
The configurations providing the best average value of the error metrics are found to require a small number of wave encounter periods of training data, between 1 and 2.
This constitutes an advantage compared to other advanced machine learning and deep learning methods, which are typically data-greedy and require computationally demanding training.

Once the model is learned, the proposed methodology offers computationally efficient prediction of ship behavior and can be exploited as an alternative to traditional high-fidelity simulations (or experiments), constituting a valuable tool for maritime design and operational planning, complementary to equation-based system identification approaches, and for real-time predictions.

The Bayesian extension to Hankel-DMDc makes the predictions more robust by incorporating uncertainty quantification and further enhances the model's predictive capabilities improving its accuracy. 
It may be noted that the Hankel-DMDc is here proposed for the prediction of ship motions from the knowledge of the initial state of a small set of variables, that can be reasonably measured on board, the rudder input, and the wave field around the vessel.
The latter has been demonstrated, \textit{e.g.}, in \cite{lee2022}, to be effectively forecastable using wave radar data.  
Coupling the proposed system identification method with such a phase-resolved wave prediction module may represent an efficient digital twinning solution for ship operations in waves.

It has been observed that some hyperparameter combinations produced spurious unstable eigenvalues in the system identification, leading to unreliable predictions.
Methodological advancements, such as the extension introduced in \cite{rains2024} which enables the possibility of enforcing the stability of the identified ROM, and the use of rank-truncated SVD in the DMDc algorithm will be explored in the future to extend the capabilities of the prediction method, overcoming some limitations that may arise in the use of the current approach in such challenging applications.

Future efforts will be devoted to considering different operational and environmental conditions, aiming to assess and improve the robustness of the system identification methods based on DMD with control, also involving transient maneuvering in waves.
In this context, the interpolation of parametric reduced-order models \cite{Farhat2008,Farhat2011} may represent an efficient methodology for extending the modeling to different sea states.
The remarkable computational efficiency of the approach may pave the way for significant advancements in the development of more accurate real-time ship simulators, incorporating realistic motion predictions and leading to safer maritime operations.

\section*{Acknowledgments}
The authors thank the financial support of the US Office of Naval Research through NICOP Grants N62909-21-1-2042 and N62909-24-1-2102, and of the Italian Ministry of University and Research through the National Recovery and Resilience Plan (PNRR), CN00000023 -- CUP B43C22000440001, “Sustainable Mobility Center” (CNMS), Spoke 3 “Waterways.” 
The authors are also grateful to NATO Science and Technology Organization, Applied Vehicle Technology task group 
AVT-351 (“Enhanced Computational Performance and Stability \& Control Prediction for NATO Military Vehicles”) for the fruitful collaboration through the years.

\section*{CRediT authorship contribution statement}
\textbf{Giorgio Palma:} Conceptualization, Methodology, Software, Validation, Investigation, Formal Analysis, Writing - Original Draft, Visualization.
\textbf{Andrea Serani:} Validation, Resources, Writing - Review \& Editing.
\textbf{Shawn Aram:} Data Curation, Resources.
\textbf{David W. Wundrow:} Data Curation, Resources.
\textbf{David Drazen:} Data Curation, Resources, Writing - Review \& Editing.
\textbf{Matteo Diez:} Conceptualization, Methodology, Resources, Writing - Review \& Editing, Supervision, Funding acquisition.


\bibliographystyle{unsrt}  
\bibliography{biblio}  

\end{document}